\documentclass[useAMS,usenatbib]{mn2e}

\usepackage{natbib}
\usepackage{graphicx}
\usepackage{rotating}
\usepackage{times}
\usepackage{amsmath}

\usepackage{bm}            
\usepackage{url}
\usepackage{color,colortbl}

\newcommand{\los}{\ensuremath{\mathrm{los}}}
\newcommand{\kms}{\ensuremath{\,\mathrm{km}\,\mathrm{s}^{-1}}}
\newcommand{\kpc}{\,\mathrm{kpc}}

\newcommand{\Msun}{\ensuremath{~\mathrm{M}_\odot}}
\newcommand{\pc}{\ensuremath{~\mathrm{pc}}}

\definecolor{Gray}{gray}{0.8}

\title[The total mass of MW]{Milky Way total Mass derived by Rotation Curve and Globular Cluster kinematics from Gaia EDR3} 

\author[Jianling Wang ]{
  Jianling Wang$^{1}$\thanks{E-mail:wjianl@bao.ac.cn},
  Francois Hammer$^{2}$,
  Yanbin Yang$^{2}$\\
{$^1$ CAS Key Laboratory of Optical Astronomy, National Astronomical Observatories, Beijing 100101, China,} \\
{$^2$ GEPI, Observatoire de Paris, CNRS, Place Jules Janssen 92195, Meudon, France.}\\
}

\begin{document} 

\date{Received ; accepted}

\maketitle 

\begin{abstract}

Using action-based distribution function for the dynamical model of the Milky
Way we have estimated its total mass and its density profile. Constraints are coming
from the globular cluster proper motions from Gaia EDR3, from the rotation
curve based on Gaia DR2 data, and from the vertical force data.
We use Bayesian Markov chain Monte Carlo method to explore the parameters, for which the globular
cluster distribution function and the Galactic potential are fully constrained.
Numerical simulations are used to study the uncertainties on the potential
constraint if considering a possible massive Large Magellanic Could (LMC). We
found that a massive LMC (1.5$\times10^{11}$ M$_{\odot}$) will affect the MW
mass measurement at large radius, which includes both the Milky Way and the
LMC. We also use the FIRE2 Latte cosmological hydrodynamic simulations to make
mock data set from a Milky-Way like galaxy that includes many unrelaxed
substructures. We test the effect of these unrelaxed substructures on the final
results, and found that the measured rotation curve fluctuated around input value
within 5 percent.  By keeping a large freedom in choosing a priori mass profile
for both baryonic and dark matter leads  a total mass of the
MW that ranges from $5.36_{-0.68}^{+0.81}\times10^{11}$ M$_{\odot}$ to
$7.84_{-1.97}^{+3.08} \times 10^{11}$ M$\odot$.  This includes the contribution
of a putative massive LMC and significantly narrows the MW total mass range
published earlier.  Such total mass leads to dark matter density at solar
position of $0.34_{-0.02}^{+0.02}$ GeV cm$^{-3}$.

\end{abstract}

\begin{keywords}
   globular clusters: general 
-- Galaxy: halo
-- Galaxy: kinematics and dynamics
-- Galaxy: structure
\end{keywords}

\section{Introduction}

The Galactic Dark Matter (DM) mass density profile and total mass are of the
most importance in modern astrophysics and cosmology.  The Milky Way (MW)
provides an unique opportunity for testing cosmology at small scales and the
galaxy formation process. The mass density profile and the total mass of the MW
governs its number of sub-halos of MW mass galaxies, which is intimately
related to low mass scale discrepancies to standard cold dark matter model
($\Lambda \mathrm{CDM}$). For example, the missing satellite and the
too-big-to-fail problems
\citep{Moore1999,Boylan-Kolchin2011,Wang2012,Cautun2014} are all closely
related to the total mass of MW. Therefore, the accurate measurement of the
total mass of MW is also important for understanding dwarf dynamics and their
accretion history, and also tests cosmological predictions. Gaia (DR2 and even
more EDR3) data become sufficiently precise to constraint the orbital
properties of MW dwarfs. It can be used to further test whether the MW and its
cortege of dwarfs are similar to $\Lambda\mathrm{CDM}$ halo and sub-haloes
\citep{Riley2019,Hammer2020,Li2021}, which again depends on the MW total mass. 

The elusive DM emits no light and can  be  only detected using indirect methods.
Since DM affects dynamics, the kinematics of various luminous
tracers have been investigated to derive its mass and density profile. In the inner
region, the disk rotation curve (RC) is usually measured with tracers having
circular motions, for instance, classic Cepheids \citep{Mroz2019}, open clusters,
HII region \citep{Sofue2012}. In the halo region, the DM mass density profile
is usually derived from kinematic analysis from halo tracers, for examples,
dwarfs assumed to be long-lived satellites \citep{Callingham2019}, globular clusters
\citep{Eadie2019,Vasiliev2019a,Watkins2019}, stellar streams
\citep{Kupper2015,Gibbons2014,Bowden2015,Malhan2019}, and halo stars
\citep{Kafle2012,Kafle2014}. A comprehensive review on the methods of
total Galactic mass measurement can be found in \citet{Wang2020}. Even though
the total mass of MW have been measured for a few decades, its actual value is still 
uncertain by a large factor \citep[see Figure 1 of][]{Wang2020}.

The $Gaia$ satellite has revolutionized the Galactic mass determination by
providing accurate proper motion for a far much larger number of stars than
ever done before. By combining the large sky spectroscopic survey in the ground
such as SDSS \citep{York2000} and LAMOST \citep{Cui2012,Zhao2012}, accurate 6D
phase-space coordinates can provide strong constraint on the Galactic DM
profile and total mass. It is expected in the following years that the $Gaia$
will continue to improve  precision and accuracy of the astrometry and
photometry for more and more stars \citep{Brown2020}. 

Understanding how the MW is structured and its assembly history is now a
central task in using those unprecedented data. Dynamical modeling is one of
most important tool to understand how the MW is structured. Dynamical modeling
with action-based distribution function DF ($f(\boldsymbol{J})$) has provided a
major progress in this field. In an axisymmetric system, the action integrals
$\boldsymbol{J}_r$, $\boldsymbol{J}_z$, $\boldsymbol{J}_\phi$  are the integral
of motions, quantifying the amplitude of oscillations in the radius and in the
vertical directions, and angular momentum around the symmetric axis,
respectively \citep{Vasiliev2019b,Binney2020}. These actions are adiabatically
invariants and conserved under slowly evolution of the potential, and in absence of energy
exchanges. Consequently, $f(\boldsymbol{J})$ is invariant too. A
system is fully determined as long as the DFs of each component are specified,
and from these DFs any measurement can be predicted for the model
\citep{Binney2020}.

Recent progresses with large spectroscopic surveys and Gaia data reveal that
the stellar halo is made of unrelaxed substructures and furthermore there might
be a large scale velocity gradient induced by the passage of the Large Magellanic
Cloud (LMC), if the latter is very massive.  These two effects may affect the
assumption of equilibrium and relaxed system in any dynamical modeling, which
should be addressed when interpretating modeling results.

In this work, we use new released data of $Gaia$ EDR3 to derive the proper
motion of Galactic globular clusters. The improvement by about a factor of 2 in
proper motion and the similar reduction of the systematic error \citep{Brown2020} make the
measurement on the MW DM density profile improvement much better than ever.
  Combining the new data with the accurate disk RC from $Gaia$ DR2
\citep{Eilers2019}, we can model the Galactic globular clusters (GCs) with the action-based
distribution function, and then constrain Galactic DM profile. By using N-body 
simulation one can test the bias introduced by a possible massive LMC. 
By using realistic cosmological hydrodynamic simulations from the FIRE2 
suite, we can test the effects of unrelaxed substructures. 

The paper is organized as it follows: Section 2 presents the measurement of GC
proper motions and their uncertainties with $Gaia$ EDR3. Section 3 describes the
additional observation data used to constrain the rotation curve (RC)
measurement.  Section 4 presents the detail on the dynamical modeling with
action-based DF method, and the results are shown in section 5. In section 6 we use
numerical simulations to investigate the possible effect of a massive LMC
passing by to the MW mass measurement, as well as the effects due to 
unrelaxed substructures. Lastly, we conclude our results in Section 7.

\section{The Proper Motion of MW GCs}

In this section we describe the method used to derive the mean proper motion 
and its associated uncertainties considering the systematic errors in the Gaia EDR3. 

\subsection{Determining the Mean Proper Motions of GCs with $Gaia$ EDR3}

We follow the procedure of \citet{Vasiliev2019a} to derive the mean proper motion (PM) and its
associated uncertainties for each cluster. We have used the publicly released 
code by \citet{Vasiliev2019a}. Here we briefly describe the method
(more details in \citet{Vasiliev2019a, Vasiliev2019c}).

The stars around each GC are clumped in PM space, which include member stars
and field stars. For each cluster, a probabilistic Gaussian mixture model is
applied to the PM distribution for stars and determine their membership
probability. A spherical Plummer profile is assumed for the prior
functional form of membership probability, for which the scale radius of
Plummer profile is allowed to be adjusted during the fitting. An isotropic
Gaussian function is assumed for the  intrinsic PM dispersion for the cluster
members.  By adopting this spatially dependent prior for the membership
probability, the intrinsic (error-deconvolved) parameters of the distributions
of both member and non-member stars are derived. Since there is non-negligible 
spatially correlated systematic errors in the PM of $Gaia$ data, this systematic 
error can be addressed by adopting the PM correlation function as below.

\begin{figure}
\center
\includegraphics[scale=0.70]{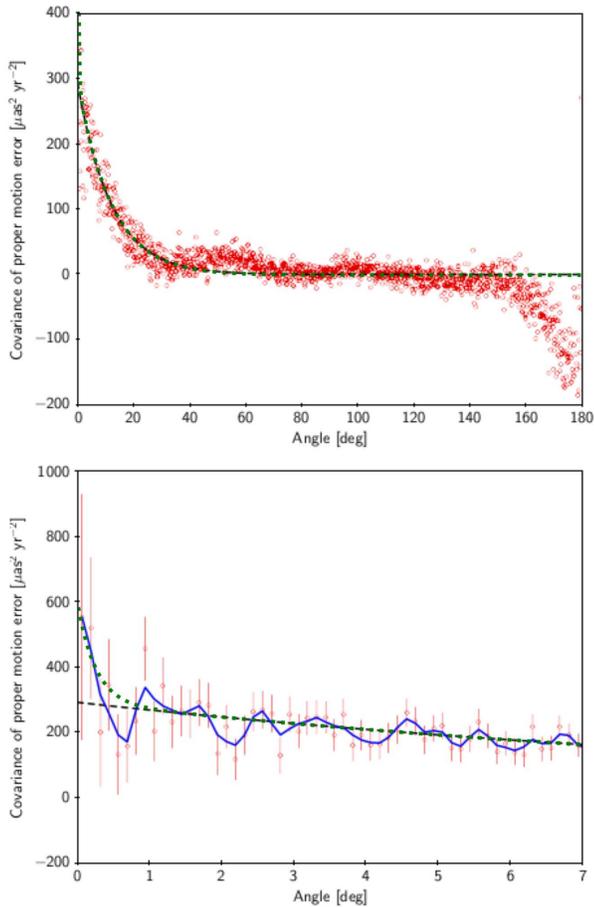}
\caption{The covariance of proper motion of data in $Gaia$ EDR3. 
This figure is from Figure 15 of  \citet{Lindegren2020}, except adding the 
new fitting curve (the dotted-green line) from eq. \ref{eq:covfnc}. The open red
circles are individual estimates and the black dash-line is an exponential
fitting. The top panel shows the large separation, while the bottom panel shows
results for small separation angle. The individual estimates and the fitted
black-dash line are from \citet{Lindegren2020}. The blue-solid line indicates 
the smoothed covariance values from \citet{Lindegren2020}. The green dashed line is our
modified fitting to the exponential fitted result as shown by eq (\ref{eq:covfnc}). }
\label{fig:Cov}
\end{figure}

\citet{Lindegren2018} and \citet{Lindegren2020} have explored the angular covariances of proper
motion based on the high precision quasar sample for both GAIA DR2 and EDR3 samples.
They have found that the covariance of proper motion errors can be well fitted
with an exponential function. This exponential function can fit well the
covariance at large scales (upper panel of Figure \ref{fig:Cov}), but it
fails to capture the variation at small scales, well below 1 degree (bottom panel of Figure
\ref{fig:Cov}).  Following \citet{Vasiliev2019c}, we modified the fitting
function of \citet{Lindegren2020} by adding another exponential function to
capture the small scale variation of covariance in the center region. 
The new function form is shown with the green dashed line in Figure \ref{fig:Cov}, 
and listed below.


\begin{equation}  \label{eq:covfnc}
\begin{aligned}
V(\theta) &= 0.000292 \exp(-\theta/12^\circ) + 0.000292\exp(-\theta/0.25^\circ) , \\
\end{aligned}
\end{equation}

$\theta$ indicates the angular separation between pairs of sources. The
first term in the right part of eq. (\ref{eq:covfnc}) is the fitted exponential
function from \citet{Lindegren2020} (the black dashed line in Figure
\ref{fig:Cov}), while the second one is our additional term to account for the
increase of covariance near the center (see the dotted-green line of Figure \ref{fig:Cov}).

Figure~\ref{fig:Cov} details how this covariance function fit to the
angular covariance of proper motion from high precision quasar from
\citet{Lindegren2020}, as well as comparing to the fitting with single
exponential function from \citet{Lindegren2020}.  The new fitting function of
eq. (\ref{eq:covfnc}) fit the covariance well for both large scale and 
central regions.

\subsection{Robustness of GC Proper Motions}

To have a clean sample with reliable astrometric measurements of the PM, we
follow the recommendations of \citet{Fabricius2020}: (1)
renormalised unit weight error (RUWE) $<1.2$. (2) asymmetric\_excess\_noise $<1.0$. (3)
ipd\_gof\_harmonic\_amplitude $<0.1$ (4) ipd\_frac\_multi\_peak $< 2$ (5) the
corrected excess factor (C$_\mathrm{corr}$) within $3\sigma$ following
\citet{Riello2020}. 

\begin{figure*}
\center
\includegraphics[scale=0.56]{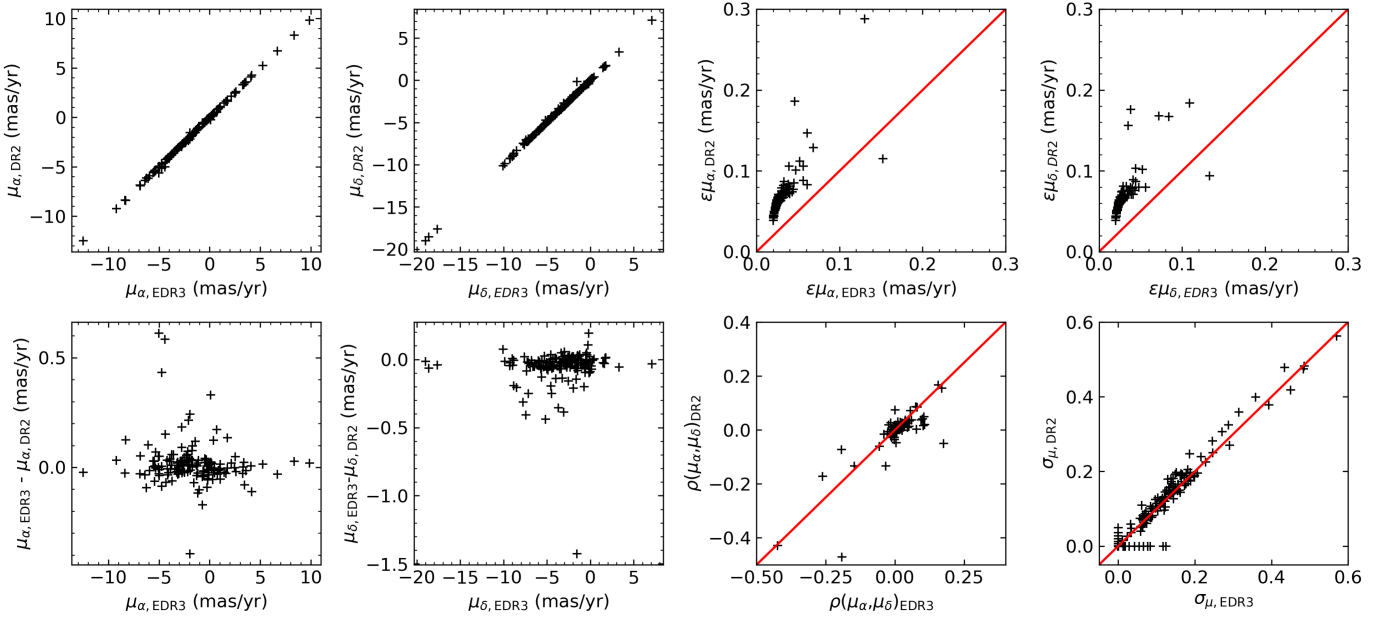}
\caption{Comparing proper motion measurement between $Gaia$ EDR3 of this work and DR2 from \citet{Vasiliev2019a}.
The panels on the top row show the one-by-one comparison of proper motion in two direction ${\mu_\alpha}, {\mu_\delta}$, and 
their errors. The left two panels in the bottom row show the difference of proper motion in two directions as function of 
proper motions. The third panel of bottom row shows the comparison of correlation coefficience of proper motion. 
The fourth panel shows the internal dispersion of proper motion (for details please refer to \citet{Vasiliev2019a}). }
\label{fig:pm}
\end{figure*}

Figure \ref{fig:pm} compare our measured proper motion, its associated
uncertainties and correlation coefficient, and the internal dispersion with
that from $Gaia$ DR2 from \citet{Vasiliev2019a}.

The first two panels in the top-left compare the proper motion of
$\mu_\alpha^*$ and $\mu_\delta$ from DR2 and EDR3 and they correlate well. In
the first two panels in the bottom-left we show the absolute differences of
proper motion as a function of $\mu_\alpha^*$ and $\mu_\delta$, respectively.
The absolute difference in proper motion is reasonably small, in most cases
lower than 0.1 mas/yr.

In the two panels in the top-right, we compare the measured uncertainties of
proper motions. The errors in EDR3 are systematic smaller than that from DR2 by
about a factor of 2, which is fully consistent with \citet{Brown2020}.

The third panel in the bottom row compares the correlated coefficients. The
coefficients are well correlated, which shows that the correlation
coefficients do not change too much even though the error of PM have decreased
a factor of 2.  The last panel of the bottom row compares the internal
dispersion. The internal dispersion are correlated between DR2 and EDR3,
which probability indicate that this method has well resolved the intrinsic
dispersion for each cluster.

\begin{figure*}
\center
\includegraphics[scale=0.5]{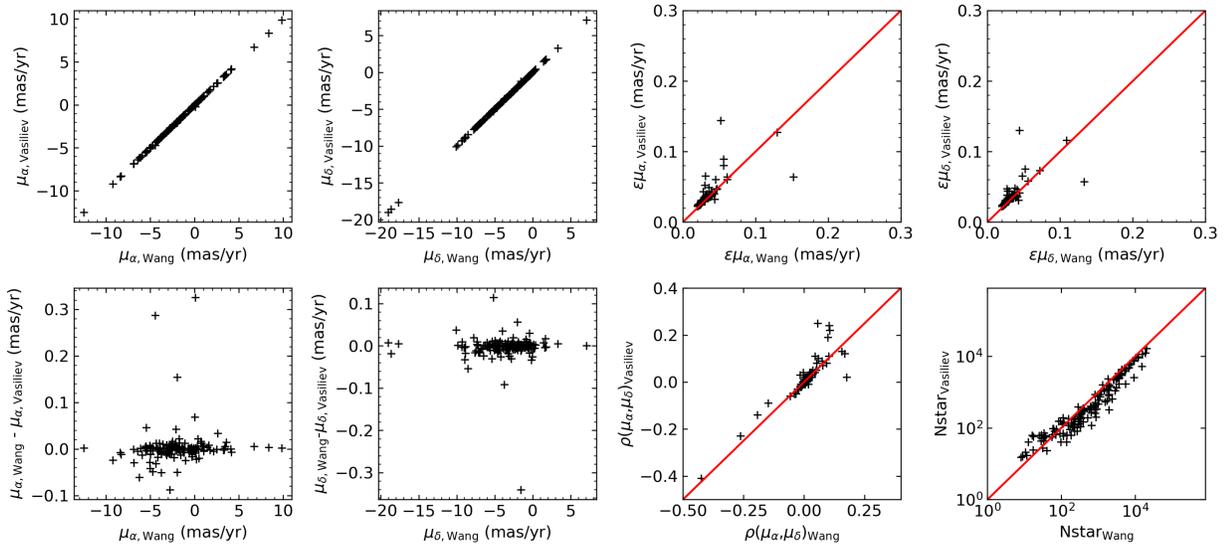}
\caption{Comparing proper motion measurements in this work with \citet{Vasiliev2021b} based on Gaia EDR3. 
The means for each panel are the same as Figure~\ref{fig:pm}. }
\label{fig:PMVasiliev}
\end{figure*}

After finishing this work, we notice \citet{Vasiliev2021b} have measured the PM
for GCs with Gaia EDR3 with an updated method comparing to \citet{Vasiliev2019a}
with a more strict selection criteria to clean the data.  We have compared their
results with ours (Figure~\ref{fig:PMVasiliev}) and find that it would not 
introduce significant differences in the final results as discussed in Appendix B.

\section{Data used to constrain MW mass modeling}

Besides GCs, here we describe other kinematic data set used to constrain the
dynamic models for Milky Way.

\subsection{Circular velocity from $Gaia$ DR2}

The rotation curve (RC) of the disk has been estimated by various tracers. 
By using the accurate proper motion from $Gaia$ DR2
combining with precise spectrophotometric parallax from APOGEE DR16,
\citet{Eilers2019} have derived disk RC for Galactocentric distance of
$5\le R\le 25$ kpc with high precision for $\sim$23000 red giants. They
have applied Jeans equation after assuming an axisymmetric gravitational
potential to obtain this measurement. This result has been found consistent
with measurement from Classic Cepheids \citep{Mroz2019,Ablimit2020}.

The precise distance and proper motion have led to the most precise RC derived to date \citep{Eilers2019}, i.e., with much smaller error bars than that of former studies.
\citet{Eilers2019} have also analyzed systematic errors from various assumption.
As mentioned by \citet{Jiao2021} that summing up all errors will dilute the significance 
of RC. Therefore, following \citet{Jiao2021} we added the systematic error of the 
cross-term in the radial and vertical velocity to the total error budgets. We noticed that 
this precise RC data have already been used to estimate MW mass \citet{Karukes2020}.

This data will provide constraints on the disk RC for our modeling. 

\subsection{Vertical force at $z=1.1$ kpc}

\citet{Piffl2014} showed that the vertical force brings an
important constraint to the dark matter shape, which has been widely used in the literature for
estimating the MW potential 
\citep{Bovy2015,Bovy2016,McMillan2017,Hattori2021}.  By using Jeans
equation \citet{Kuijken1991} measured vertical force, $K_{z,1.1 \mathrm{kpc}}$
at $z=1.1$ kpc away from the disk.  \citet{Bovy2013} using SDSS/SEGUE data
measured the vertical force $K_{z,1.1 \mathrm{kpc}} (R)$ as function of radius,
which is used in our analysis. As mentioned by \citet{Hattori2021} the RC of
\citet{Eilers2019} provides the radial force constraint on the model, while the
$K_{z,1.1 \mathrm{kpc}} (R)$ gives an independent constraint on the vertical
direction.

\section{Modeling}

In the following sections, we describe how the modeling for the GC DF
has been setup.  We note the six GCs associated with Sagittarius galaxy (NGC 6715,
Terzan 7, Terzan 8, Arp 2, Pal 12, and Whiting 1), which we have excluded in the analysis
following  \citet{Vasiliev2019a} and \citet{Myeong2019} and since they are clustering 
in the phase-space diagram. There could be other GCs associated 
with different accretion events \citep{Massari2019} and they are included in current 
analysis. However, in section \ref{sec:unrelaxed} we verify that unrelaxed substructures do not 
have significant effect on the final results, on the basis of an analysis made using the FIRE2 suite of simulations. 

\subsection{Models for the galactic gravitational potential}

In this work we use axisymmetric Galactic potential $\Phi(R,z)$, which consist
of baryon mass components and dark matter.  In order to determine the dark
matter mass profile, we first specify the adopted baryon mass model that follows
\citet{McMillan2017}, and which include multi-components (a bulge, a thin and a thick
stellar disc, an H I disc and a molecular gas disc).

\citet{Jiao2021} investigated different baryonic mass profiles to test the MW
dark matter distribution. Here we prefer to adopt a baryonic mass profile that
also includes the gas contribution. Given the fact that baryons are not the
dominant component, this should not alter our main conclusions.

\subsubsection{The bulge}

The stellar bulge component is modeled as \citet{McMillan2017} as following:

\begin{equation}\label{eq:bulge}
   \rho_\mathrm{b}=\frac{\rho_{0,\mathrm{b}}}{(1+\frac{r^\prime}{r_0})^{\alpha}}\;
  \textrm{exp}\left[-(\frac{r^\prime}{r_\mathrm{cut}})^2\right],
\end{equation}

where, in cylindrical coordinates,
\begin{equation}
  r^\prime = \sqrt{R^2 + (\frac{z}{q})^2}
\end{equation}

There is a bar in the central region of the MW, which introduces non-axisymmetric potential. 
In this work, we use the software $AGAMA$ \citep{Vasiliev2019b} to deal with actions with St\"ackel fudge method, which 
can only handle oblate axisymmetric potentials. Therefore, the central region is not expected to be 
well modelled. Following the literature \citep{Vasiliev2019a, Hattori2021, Cautun2020}, 
the parameters $r_0, \alpha, r_{\mathrm{cut}}$, and $a$ parameters are fixed during 
the modeling procedure, and their values are listed in Table \ref{tab:FixPars}. 
Following \citet{Cautun2020}, the scale density of $\rho_{0,\mathrm{bulge}}$ has a Gaussian prior value 
of $100\pm 10$ M$_{\odot}$ pc$^{-3}$.

\subsubsection{The thin and thick stellar disk}

The stellar disk is modeled by a thin and a thick disk component, which are both described by the following 
exponential profile:

\begin{equation} \label{eq:MW_disc} 
 \rho_{\mathrm{d}} (R,z) = \frac{\Sigma_{0}}{2z_{\mathrm{d}}} \; \exp\left( -\frac{\mid z\mid}{z_{\mathrm{d}}} - \frac{R}{R_{\mathrm{d}}} \right) \;,
\end{equation}

Here $z_d$ denote the scale-height for the disk. Following \citet{McMillan2017}
we set $z_d$ to be 0.3 and 0.9 kpc for thin and thick disk, and keep fixed
during our modeling process. $\Sigma_{0}$ and $R_\mathrm{d}$ indicate the
center surface density and scale-length for the disks. These two parameters for
each disk components are not fixed, but fitted with Gaussian prior values.
These prior values are from \citet{McMillan2017}, except for the
thick disk scale-length ($R_\mathrm{d}$) ,for which we adopt a value from 3.5 kpc from \citet{Bland-Hawthorn2016}
and \citet{Cautun2020}.  A 30\% uncertainty is adopted for the $\Sigma_0$
and $R_\mathrm{d}$ during the fitting.

\subsubsection{The molecular and atomic gas discs}

The atomic (H$_\mathrm{I}$) and molecular (H$_2$) gas disk parameters and form
are adopted from \citet{McMillan2017}, and:

\begin{equation}
    \rho_{\mathrm{g}} (R,z) = \frac{\Sigma_{0}}{4z_{\mathrm{d}}} \;
    \exp\left( -\frac{R_{\rm hole}}{R} - \frac{R}{R_\mathrm{d}} \right) \;
    {\rm sech}^2\left( \frac{z}{2z_{\mathrm{d}}} \right)
    \label{eq:gas} \;,
\end{equation}

These models are kept fix during the
procedure.  In the model, the mass of H$_\mathrm{I}$ is 1.1$\times 10^{10}$
M$_{\odot}$, and the molecular gas mass is around 10\% of the H$_\mathrm{I}$.

\subsubsection{The dark matter halo}

We chose two well-known and used mass profiles to get flexible the slope of
DM density at outskirts, one follows \citet[hereafter called 'Zhao']{Zhao1996}
and the other \citet[hereafter called 'Einasto']{Einasto1965}.

There are five parameters in the Zhao's profile:

\begin{equation}  \label{eq:Zhao}
\rho(r) = \rho_\mathrm{0,h}\,\left(\frac{r}{r_\mathrm{h}}\right)^{-\gamma}
\left[ 1 + \left(\frac{r}{r_\mathrm{h}}\right)^\alpha \right] ^ { (\gamma-\beta) / \alpha }.
\end{equation}

In the case of $\alpha=1, \beta=3$, and $\gamma=1$ the profile corresponds to the
NFW profile, which is widely used in the literature for representing dark
matter halos.  Comparing to the NFW profile, the Zhao's profile has more
flexibility. 

For the Einasto profile, we adopt:

\begin{equation} \label{eq:Einasto}
\rho(r) = \rho_\mathrm{0,h}\,\exp\left[  \left(-\frac{r}{r_\mathrm{h}}\right)^{-\beta} \right]
\end{equation}

There are three parameters in the Einasto profile, which has
been argued to provide the best description of the dark matter profile
\citep{Gao2008,Bullock2017}. 

The determination of Galactic DM shape is important to constraint
cosmological models.  The shape of DM halo is triaxial \citep{Jing2002} in
DM-only simulation, and this can be affected by the inclusion of baryonic
component. \citet{Chua2019} using Illustris suite of simulation found that the
DM halo has an oblate-axisymmetric shape with a minor to major ratio of $0.75\pm
0.15$.  The MW dark halo shape has been measured with various methods.
\citet{Wegg2019} used RR Lyrae stars and they found
the flattening of DM halo $q=1.00\pm0.09$, based on the axisymmetric Jeans equation.  The GD-1 stream kinematics has been
used to measured the Galactic DM shape, and \citet{Malhan2019} found
$q=0.82\pm^{+0.25}_{-0.13}$, while \citet{Bovy2016} gave $q=1.3\pm^{+0.5}_{-0.3}$.
\citet{Vasiliev2021a} modeled  the Sagittarius stream considering the effect of a massive LMC
on the MW using an oblate DM halo, which becomes triaxial beyond 50 kpc.
With axisymmetric Jeans equations, \citet{Loebman2014} considered SDSS halo stars 
and estimated the MW DM density flattening to be $q=0.4\pm0.1$. Therefore, the large range
of $q$ from different observations leads us to use a  large range of flattening parameters for 
the modeling.

During the modeling, the halo shape is not spherical, and we have let the halo
shape parameter $q$ free varying from 0.1 to 1, which corresponds to an oblate
shape ($q\le1$). In current studies, we are limited to oblate halos, which
is a restriction due to the AGAMA software \citep{Vasiliev2019b}. The lower limit
is set to avoid calculation divergence.

All of the parameters for the two dark matter halo profiles are free in our
modeling. In Table \ref{tab:FitPars} all free parameters of the
modeling are listed. 
Even though halo parameters are limited to ranges listed in Table \ref{tab:FitPars}, 
we have checked the MCMC chains for each parameter to be sure that  
parameters have been explored in sufficiently large ranges, to ensure the absence of non-investigated solutions. 

\subsection{Distribution function}

By assuming that the GC system is in dynamical equilibrium, the distribution function
(DF) of GCs can be expressed in phase-space by a function of
$f(\boldsymbol{J})$ with three actions, $\boldsymbol{J}=(J_r, J_z, J_\phi)$,
where $J_r$ and $J_z$ is the radial and vertical actions, and $J_\phi$ is the
azimuthal action and equal to angular momentum in the $z$ component.

There are evidences for two distinct GC populations, one being
metal-rich and the other metal-poor and the former show rapid
rotations and are concentrated in the center \citep{Harris1979,Zinn1985}.  Unlike
\citet{Binney2017} and \citet{Posti2019} using two components in the DF to
model the distribution function of GCs, we use a double-power-law DF
\citep{Vasiliev2019a}.  As shown in \citet{Vasiliev2019a} this DF is flexible
enough to describe both populations reasonably well, and the gain is to have less free
parameters. 

\begin{equation}  \label{eq:DF}
\begin{aligned}
f(\boldsymbol{J}) &= \frac{M}{(2\pi\, J_0)^3}
\left[1 + \left(\frac{J_0}{h(\boldsymbol{J})}\right)^\eta \right]^{\Gamma/\eta}  
\left[1 + \left(\frac{g(\boldsymbol{J})}{J_0}\right)^\eta \right]^{-\mathrm{B}/\eta} 
\makebox[-1cm]{}\\
&\times \bigg(1 + \tanh\frac{\kappa J_\phi}{J_r + J_z + |J_\phi|}\bigg). 
\end{aligned}
\end{equation}

where

\begin{equation*}
\begin{aligned}
g(\boldsymbol{J}) &\equiv g_r J_r + g_z J_z\, + (3-g_r-g_z)\, |J_\phi|, \\
h(\boldsymbol{J}) &\equiv h_r J_r + h_z J_z   + (3-h_r-h_z)   |J_\phi|
\end{aligned}
\end{equation*}

The dimensionless parameters $(g_r, g_z, g_\phi)$ and $(h_r, h_z, h_\phi)$
control the density shape and the velocity ellipsoid in the outer region and inner
region \citep{Posti2015, Das2016a, Das2016b,Vasiliev2019b}, respectively. $g_i$
and $h_i$ have been constrained by $\Sigma_i h_i=\Sigma_i
g_i=3$\citep{Vasiliev2019b}. In this way the degeneracy between $g_i$ and $h_i$
and $\boldsymbol{J}_0$ will be broken\citep{Das2016a, Das2016b}.  The power-law
indices $\mathrm{B}$ and $\mathrm{\Gamma}$ are related to the outer and inner
slope, while $\eta$ determine the steepness of this two regime transition.  The
parameter $\kappa$ controls the net rotation of the system, with $\kappa=0$
being the non-rotation case, and $\kappa={\pm}1$ indicate the maximal
rotation case. In the publicly released version of AGAMA
\citep{Vasiliev2019b}, $\boldsymbol{J}_\phi$ is normalized by a fixed constant,
which leads to a non-rotating core. Following \citet{Vasiliev2019a} we have
modified the  publicly released software AGAMA, in the way that the
$\boldsymbol{J}_\phi$ is normalized by the value summarizing three actions and
the rotation will be roughly constant at all energies. The total mass is the
normalization parameter.

\begin{table}
\caption{Parameters of baryon gravitational potential are fixed in the dynamical model.} 
\label{tab:FixPars}
\centering
\begin{tabular}{llllll}
\hline
\rowcolor{Gray}
$\quad$         &
H$_{\mathrm I}$ disc   $\;$ &
H$_2$ disc   $\;$      &
Units       $\;$      &\\
\hline
$\Sigma_0$         & 53.1 &2179.5 & $M_\odot\,\textup{pc}^{-2}$\\
$R_\textup{d}$     & 7.00 & 1.5   & kpc\\
$z_\textup{d}$     & 0.85 &0.045  & kpc\\
$R_\textup{hole}$  & 4    & 12    & kpc\\
\hline
\rowcolor{Gray}
$\quad$         &
Stellar thin  disc   $\;$    &
Stellar thick disc   $\;$    &
Units       $\;$     & \\
\hline
$z_\textup{d}$    & 0.3 & 0.9 & kpc \\
\hline
\rowcolor{Gray}
$\quad$            &
Stellar bulge $\;$ &
Units       $\;$      &
$\quad$     $\;$     &\\
\hline
$r_0$            & 0.075  & kpc                        \\
$r_\textup{cut}$ & 2.1    & kpc                        \\
$\alpha$         & 1.8                                 \\
$q$              & 0.5                                 \\
\hline
\end{tabular}
\end{table}

\subsection{Error models for observables}\label{sec:ErrModel}

Observations of the GC system are not error-free. In the
following we consider Gaussian models for the error  to associate the true quantities with
observables and its errors. The  six observables for GCs are $\bar{u}$
$= (l, b, s, v_{\los}, \mu_\alpha^*, \mu_\delta)$, where $(l,b)$ denote the
Galactic longitude and latitude which are measured with high precision so their
errors are neglected in the following analysis. The heliocentric distance is $s$ and its error is not neglected. Following \citet{Vasiliev2019a} we
adopted a 0.046 percent uncertainty (correspond to 0.1 in distance modulus). 
$v_\los$ is the line-of-sight velocity, and its value is taken from the
Table C1 of \citet{Vasiliev2019a}. The proper motion is derived from the above
study, and the correlated uncertainties in
$\bf{\mu}=(\mu_\alpha^*, \mu_\delta)$ as well as the covariance matrix
$(\Sigma_{\mu})$ are taken into account.

In the following, a Gaussian function is used to associate the observables
$(\bar{u})$ to their true value $(u)$. The error models for the heliocentric
distance and line-of-sight velocity are:

\begin{equation}
N(s |\bar s, \sigma_s) =  \frac{1}{\sqrt{2\pi\sigma_s^2}} \exp \left[ - \frac{(\bar{s} - s)^2}{2\sigma_s^2} \right],
\end{equation}

\begin{equation}
N(v_\los |\bar v_\los, \sigma_{v_\los}) =  \frac{1}{\sqrt{2\pi\sigma_{v_\los}^2}} \exp \left[ - \frac{(\bar{v} - v)^2}{2\sigma_{v_\los}^2} \right],
\end{equation}

For the proper motions:

\begin{equation} \label{eq:pmErr} 
N(\bf{\mu} | \bf{\bar{\mu}}, \Sigma_{\mu}) = \frac{1}{ 2\pi |\Sigma_\mu|^{1/2}} \exp \left[ - \frac{1}{2} (\bf{\mu}-\bf{\bar{\mu}})^\mathrm{T} \Sigma_\mu^{-1} (\bf{\mu}-\bf{\bar{\mu}}) \right]
\end{equation}

\subsection{The Bayesian inference}

With the Bayes theorem, we can determine the posterior distribution of the model
parameters $(M)$ given the data $(D)$. From this posterior distribution, the 
model parameters and their credible regions are estimated.

\begin{equation} \label{eq:Baye}
\mathrm{Pr}(M|D ) = \frac{\mathrm{Pr}(D|M) \times \mathrm{Pr}(M)}{\mathrm{Pr}(D)}
\end{equation}
where $\mathrm{Pr}(D|M)$ is the likelihood of the data given the model
parameters, $\mathrm{Pr}(M)$ is the prior probability ascribed to the set of
parameters, and $\mathrm{Pr}(D)$ is a normalization factor. In the following, we
show how the total likelihood is built from the model and the priors.

\subsubsection{Likelihood for the GC distribution function}

\begin{scriptsize}
\begin{equation}
\begin{aligned}
\ln \mathcal{L}_\mathrm{GCs} = \sum_{i=1}^{N_\mathrm{clusters}} \ln 
\frac
{
S(\bar{\boldsymbol{u}}_i)
\int \mathrm{d}^6 \boldsymbol{u} \;
E( \bar{\boldsymbol{u}}_i |  \boldsymbol{u} , M )
f( \boldsymbol{u} | M)
\left|
\frac
{\partial (\boldsymbol{x}, \boldsymbol{v})}
{\partial \boldsymbol{u}}
\right|
}
{
\int \mathrm{d}^6 \boldsymbol{u}'
\int \mathrm{d}^6 \boldsymbol{u} \;
E( \boldsymbol{u}' |  \boldsymbol{u} , M )
f( \boldsymbol{u} | M)
S(\boldsymbol{u}')
\left|
\frac
{\partial (\boldsymbol{x}, \boldsymbol{v})}
{\partial \boldsymbol{u}}
\right|
} .
\end{aligned}
\end{equation}
\end{scriptsize}

where ${{\boldsymbol{u}'}}$ indicates the true value of the observational
vector.  $\mathrm{E(\bar{\boldsymbol{u}} |  \boldsymbol{u} , M)}$ denotes
the error model (see section \ref{sec:ErrModel}) for the probability of observables
$(\mathrm{\bar{\boldsymbol{u}}})$, given a model $M$ and the true values
$\mathrm{\boldsymbol{u}}$.  $f(\boldsymbol{u} | M)$ indicates the probability
that a GC has a true vector ${\boldsymbol{u}}$ given a model $M$. $\left|\frac
{\partial (\boldsymbol{x}, \boldsymbol{v})} {\partial \boldsymbol{u}}\right|$
is the Jacobian factor for transformation of coordinate system with value
$s^4$cos$\delta$.  Following \citet{Posti2019} and \citet{Vasiliev2019a} we
have neglected the selection function ($S({\boldsymbol{u}}$)=1) on the GCs, since it has
little effect on the model parameter inference, as
demonstrated by \citet{Binney2017}. In this case the integration in
the denominator is the normalization factor, which is the total number of GCs
and that is identical for each cluster. The integral in the numerator is calculated
with Monte Carlo sampling technique with fixed sampling points and
a weighting value to reduce the noise
\citep{Binney2017,Vasiliev2019a,Hattori2021,Das2016a,Das2016b,McMillan2013}.

\subsubsection{Likelihood from the disk circular data}

The precise rotation curves in the disk region have been derived by
\citet{Eilers2019}, which can provide constraints on the total potential and can
help to break the degeneracy between baryonic and DM contributions to the potential.
This motivate us to include  rotation curve data into our modeling with
te Bayesian theorem.

For a given set of model parameters, the circular rotation curve for a given
radius $R$  at the meridian plane($z=0$), can be derived from the following equation:

\begin{equation}
v_\mathrm{circle}^\mathrm{model} (R) = \left[
R \left(
\frac{\partial \Phi (R,z=0)}{\partial R}
\right)
\right]^{1/2}.
\end{equation}

Following \citet{Hattori2021}, the sum of the logarithm of the likelihood for the observed rotation 
curve from \citet{Eilers2019} can be derived as :

\begin{equation} \label{eq:LK_RC} 
\ln \mathcal{L}_\mathrm{circle} 
= - \sum_{i=1}^{N_\mathrm{circle}} \sqrt{2\pi}\sigma_i  + \frac{1}{2} \left( 
\frac{v_\mathrm{circle}(R_{i}) - v_\mathrm{circle}^\mathrm{model} (R_{i})}
{\sigma_i} \right)^2 .
\end{equation}

$v_\mathrm{circle}^\mathrm{model} (R_{i})$ indicates the rotation curve at
each radial position of the observation data as done by \citet{Eilers2019}.
$v_\mathrm{circle}(R_{i})$ and $\sigma_i$ give the observed rotation curve
and its associated uncertainties of measured rotation curve at different radius
from \citet{Eilers2019} and \citet{Jiao2021}.

\subsubsection{Likelihood for the vertical force $K_{z, 1.1 \mathrm{kpc}}$}

For a given set of model parameters, the vertical force at position of $(R, z=\mathrm{1.1 kpc})$
can be derived as it follows:

\begin{equation}
K_{z=1.1 \kpc}^\mathrm{model} (R) =
\left[
- \frac{\partial \Phi (R,z=1.1 \kpc)}{\partial z}
\right] .
\end{equation}

Using the G dwarfs data from  SDSS/SEGUE survey \citep{Lee2011},
\citet{Bovy2013} derived the vertical force at $z=1.1$ kpc
($K_{\mathrm{z=1.1}}(R_i)$) at several radii assuming different 'mono-abundance'
population. By requiring that the spatial distribution and
the vertical kinematics are consistent with the phase-space data of observations,
they derived an independent gravitational potential and a three-integral 
action-based DF for each sub-population, from which they derived the $K_z$ at 
different radii. 
With the observation data for the
vertical force at $z=1.1$ kpc ($K_{\mathrm{z=1.1}}(R_i)$) at different radii
$R_i$ and their associated errors ($\sigma_\mathrm{Kz}( R_{i} )$) from
\citet{Bovy2013}, the sum of logarithmic likelihood is derived from the
following:

\begin{scriptsize}
\begin{equation} \label{eq:LK_Kz}
\ln \mathcal{L}_\mathrm{Kz}  
= - \sum_{i=1}^{N_\mathrm{Kz}} \sqrt{2\pi}\sigma_\mathrm{Kz}( R_{i}) + \frac{1}{2} \left( 
\frac{K_\mathrm{z=1.1 kpc}( R_{i} ) - K_\mathrm{z=1.1 kpc}^\mathrm{model} ( R_{i} )}
{\sigma_\mathrm{Kz}( R_{i} )} \right)^2 . 
\end{equation}
\end{scriptsize}

\subsubsection{Total likelihood}

We take a simple and reasonable assumption that given the model
parameters, the above three observation data set are conditionally independent,
which do not provide additional information about each
other. Then from the above derivation, the total logarithmic likelihood for a
given set model parameters can be expressed as:

\begin{equation}
\ln \mathrm{Pr}(D|M) = \ln \mathcal{L}_\mathrm{GCs} + \ln \mathcal{L}_\mathrm{Kz} + \ln \mathcal{L}_\mathrm{circle}
\end{equation}

\subsubsection{The Priors}

In the Bayesian inference, we can put priors to constrain the amplitude of
 parameters.  Our priors are listed in Table \ref{tab:FitPars}.  The prior in
the baryon gravitational potential is mostly taken from \citet{McMillan2017},
\citet{Deason2021}, and \citet{Bland-Hawthorn2016}, and a Gaussian function is adopted for the prior
function.  For the parameters related to the dark matter profile and DF
of GCs, the priors are set as uniform within the a reasonable ranges listed in
Table \ref{tab:FitPars}. In the Cold Dark Matter (CDM), the dark matter
halo follow a cuspy density profile with $\gamma \sim 1$ (NFW), however, the
observed rotation curve of local spirals seems to be more consistent with core
density profile with $\gamma \sim 0$. The core density profile is also reasonable for
dwarf spheroidal galaxies and Low Surface Brightness (LSB). (see \citealt{Matteo2008} for
discussion). The situation become more complex if the dark matter profile
is modified after the inclusion of baryons \citep{Cautun2020}, which results
in the fact that neither NFW nor the generalized NFWsucceeded to fit the MW rotation curve data. However
\citet{Jiao2021} found that a nearly flat density core with Einasto profile is best
for MW dark matter density profile, including baryons or not. Based on
the above discussion, we decide to adopt non-informative flat priors for the dark
matter profile parameters.  
For the parameters relevant to the DF of GCs, we chose uniform priors following 
the literature \citep{Vasiliev2019a,Posti2019,Binney2017}.
We have visually checked the posterior 
distribution for the MCMC chains to be sure that the prior range is large 
enough and do not impose constraints on the parameters sampling.

\subsection{Model parameter estimates}

We use the Nelder-Mead method implemented in the python scipy package to
maximize the above likelihood, and find the parameters with maximum-likelihood.
By using these parameters as initial input values, we use Monte Carlo Markov
Chain (MCMC) method to explore the parameter space, which is implemented in the
EMCEE package \citep{Foreman-Mackey2013}. To be sure a converged results
achieved with MCMC, we run $\sim 10\times$ N$_\mathrm{pars}$ (where N$_\mathrm{pars}$
is the total number of free parameters) walkers for the modeling of the GC system,
and $\sim 5\times$ N$_\mathrm{pars}$ for the mock simulation data in section 6.
The MCMC is ran for several thousand steps to be sure to achieve a
converging result, and in the following analysis, the first half chain is
discarded for the initial burn-in chain. We use the median value of the
posterior distribution for the estimated results, and 68 percentile for the
credible intervals. We point out that 68 percentile does not reflect a one $\sigma$ error bar since the marginal posterior distribution is non-Gaussian.

\section{Results on the MW mass}


\subsection{The posterior distribution of parameters}

\begin{figure*} 
\center
\includegraphics[scale=1.20]{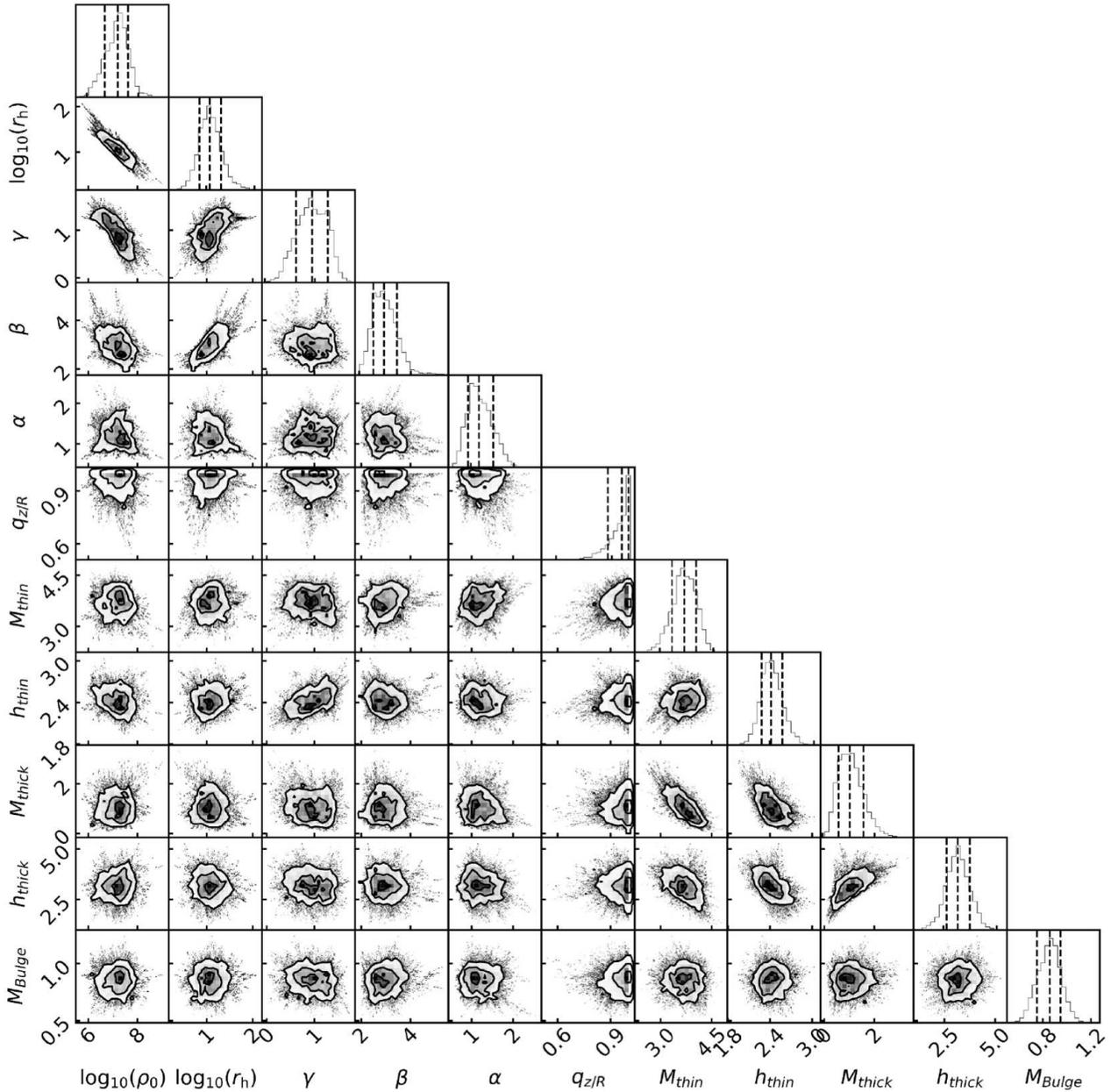}
\vspace*{5mm}
\caption{Posterior distribution of parameters for potential fields with model
using the Zhao's DM density profile (Eq.\ref{eq:Zhao}). The parameters $\log
\rho_0, \log a_\mathrm{scale}, \gamma, \beta$ and $\alpha$, are for parameters of
dark matter mass distribution. The parameters $M_\mathrm{thin},
M_\mathrm{thick}, M_\mathrm{bulge}$ are total mass for the thin and thick disk,
and bulge components, with units in $10^{10}$ M$_{\odot}$.  The parameters $h_\mathrm{thin}, h_\mathrm{thick}$
indicate the scale-length for thin and thick disk, respectively.  Contour lines in each panel and 
the vertical lines in the marginal histograms show the 16\%, 50\%, 84\% percentiles.}
\label{fig:Pot}
\end{figure*}

\begin{figure*} 
\center
\includegraphics[scale=1.20]{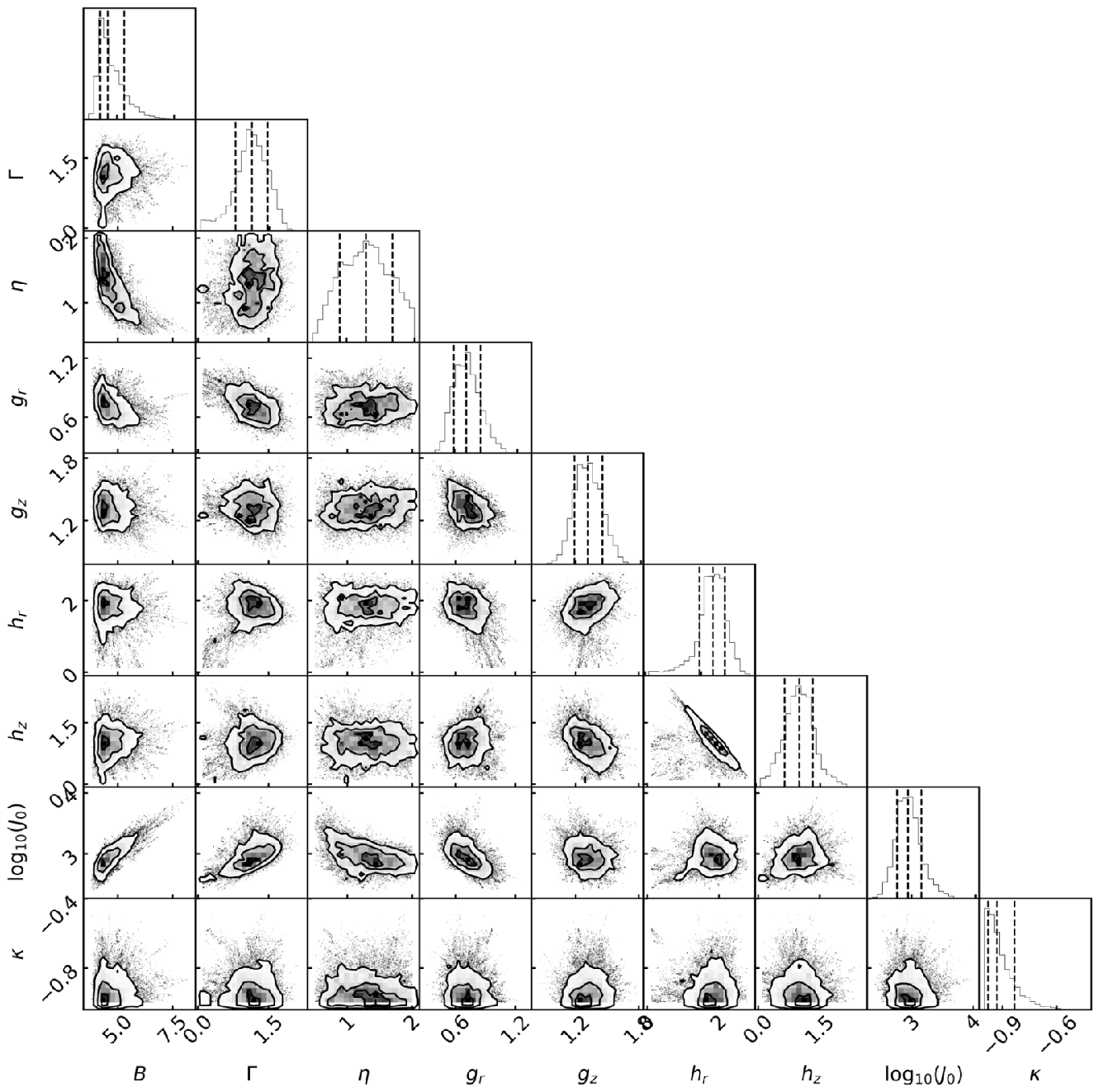}
\vspace*{5mm}
\caption{Posterior distribution of parameters of DF (eq. \ref{eq:DF}) of GCs
with model using the Zhao's DM density profile (Eq.\ref{eq:Zhao}).  
The parameters $B$ and $\Gamma$ related to the outer and inner slopes of
DF, while $\eta$ determine the steepness of transition. The dimensionless
parameters $g_r, g_z, h_r, h_z$ control the density shape and the velocity
ellipsoid in the outer and inner region. The inner and outer regions are separated 
with actions $\mathrm{J_0}$. $\kappa$ that control the rotation. Contour lines in each panel and
the vertical lines in the marginal histograms show the 16\%, 50\%, 84\% percentiles. 
}
\label{fig:DF} 
\end{figure*}

In this Section we examine the final results from analyzing the posterior
distribution. To have an overall view about the estimated parameters of our
modeling, we show the posterior distribution of the inferred parameters with Zhao
DM density profile (Eq.\ref{eq:Zhao}) in Figure~\ref{fig:Pot} and \ref{fig:DF}.
The posterior distribution of estimated parameters are separated into
gravitational potential fields (Figure~\ref{fig:Pot} ) and  distribution
function of GCs (Figure~\ref{fig:DF}), respectively. The posterior distribution
of parameters show that they converge well.  The final results are listed in
Table \ref{tab:FitPars} in the last columns, with both the DM density profiles for
Einasto (Eq.\ref{eq:Einasto}) and Zhao (Eq.\ref{eq:Zhao}) models. 

\begin{table*}     
\centering
\caption{The model parameters used in our modeling. The best derived values
are shown with median and 68 percentile of the posterior distribution. Gaussian prior 
functions have been used for baryon model parameters, while flat prior has been adopted 
for all the other parameters. The ranges of parameters in the prior have been chosen which 
are large enough without imposing constraints on the parameters sampling with MCMC when checking the MCMC chains. 
The low limit of out slope ($\beta$) for DM profile in Zhao is set to 2, and 0 for 
Einasto profile.}
    \renewcommand{\arraystretch}{1.4}
    \begin{tabular}{ p{3.2cm} p{1.7cm}p{2.0cm} lrr} 
        \hline\hline
        Parameters & Symbol & Units & Prior & \multicolumn{2}{c}{Best fitting values} \\[-.2cm]
        & & & &  Einasto halo & Zhao's halo \\
        \hline\\[-.4cm]        
        \multicolumn{6}{c}{\bf Gravitational Potential} \\
        \hline

        \rowcolor{Gray}
        \multicolumn{6}{c}{\it Baryon gravitational Potential} \\
        bulge density           & $\rho_{0,\rm bulge}$   & $\Msun\pc^{-3}$& $100\pm10$ & $94.64_{-9.76}^{+9.35}$      & $95.20_{-11.79}^{+9.90}$       \\
        thin disc density       & $\Sigma_{0,\rm thin}$  & $\Msun\pc^{-2}$& 900$\pm$270& $1057.50_{-89.42}^{+87.71}$  & $1003.12_{-130.70}^{+134.77}$ \\ 
        thick disc density      & $\Sigma_{0,\rm thick}$ & $\Msun\pc^{-2}$&183$\pm$55  & $167.76_{-54.09}^{+57.40}$   & $167.93_{-54.06}^{+60.09}$  \\ 
        thin disc scale length  & $R_{\rm thin}$         & $\kpc$         & $2.5\pm0.5$& $2.39_{-0.11}^{+0.11}$       & $2.42_{-0.13}^{+0.15}$   \\ 
        thick disc scale length & $R_{\rm thick}$        & $\kpc$         & $3.5\pm0.7$& $3.20_{-0.54}^{+0.52}$       & $3.17_{-0.54}^{+0.56}$ \\ 
        \rowcolor{Gray}

        \multicolumn{6}{c}{\it DM density profile} \\
        DM density         & $\rho_0$  & $\Msun\kpc^{-3}$& $ 0<\log_{10} \rho_0<15$ & $9.29_{-0.21}^{+0.20}$  &  $ 7.19_{-0.51}^{+0.38}$    \\ 
        DM scale length    & $r_h$     & $\kpc$          & $-2<\log_{10} r_h<4.5$   & $-1.40_{-0.26}^{+0.26}$ &  $1.07_{-0.21}^{+0.24}$  \\ 
        Inner slope        & $\gamma$  &  --             & $0<\gamma<3$             &     --                  &  $0.95_{-0.32}^{+0.31}$   \\ 
        steepness          & $\alpha$  &  --             & $0 < \alpha<20$          &     --                  &  $1.19_{-0.25}^{+0.33}$  \\ 
        Outer slope        & $\beta$   &  --             & $2^\mathrm{Zhao},0^\mathrm{Eina}<\beta<20$   & $0.32_{-0.02}^{+0.02}$  &  $2.95_{-0.41}^{+0.51}$   \\ 
        axis ratio $(z/R)$ & $q$       &  --             & $0.1<q<1$                & $0.97_{-0.06}^{+0.03}$  &  $0.95_{-0.07}^{+0.04}$   \\ 


        \hline \\[-.4cm]
        \multicolumn{6}{c}{\bf Distribution function of GCs} \\
        slopeOut       & $B$           & --     &$3.2<B<10$        &   $5.03_{-0.64}^{+1.88}$     & $4.61_{-0.35}^{+0.71}$     \\ 
        slopeIn        & $\Gamma$      & --     &$0.1<\Gamma<2.8$  &   $1.23_{-0.28}^{+0.26}$     & $1.14_{-0.34}^{+0.32}$     \\ 
        steepness      & $\eta$        & --     &$0.5<\eta<2.0$    &   $1.08_{-0.37}^{+0.52}$     & $1.29_{-0.39}^{+0.39}$     \\ 
        coefJrOut      & $g_r$         & --     &$0.1<g_r<2.8$     &   $0.65_{-0.13}^{+0.12}$     & $0.71_{-0.12}^{+0.14}$     \\ 
        coefJzOut      & $g_z$         & --     &$0.1<g_z<2.8$     &   $1.32_{-0.13}^{+0.11}$     & $1.32_{-0.12}^{+0.13}$     \\ 
        coefJrIn       & $h_r$         & --     &$0.1<h_r<2.8$     &   $1.86_{-0.29}^{+0.31}$     & $1.81_{-0.37}^{+0.32}$      \\ 
        coefJzIn       & $h_z$         & --     &$0.1<h_r<2.8$     &   $1.01_{-0.30}^{+0.29}$     & $1.01_{-0.34}^{+0.32}$     \\ 
        J$_0$          & $J_0$         & --     &$-2<\log_{J_0}<7$ &   $3.08_{-0.22}^{+0.45}$     & $2.94_{-0.18}^{+0.21}$     \\ 
        rotFrac        & $\kappa$      & --     &$-1<\kappa<1$     &   $-0.94_{-0.04}^{+0.08}$    & $-0.93_{-0.05}^{+0.10}$    \\

        \hline \\[-.4cm]
        \multicolumn{6}{c}{\bf Derived quantities} \\
        bulge mass         & $M_{\star,\rm bulge}$ & $10^{10} \Msun$ &--&  $0.85_{-0.08}^{+0.09}$   &    $0.86_{-0.11}^{+0.09}$   \\ 
        thin disc mass     & $M_{\star,\rm thin}$  & $10^{10} \Msun$ &--&  $3.79_{-0.30}^{+0.28}$   &    $3.69_{-0.37}^{+0.34}$   \\ 
        thick disc mass    & $M_{\star,\rm thick}$ & $10^{10} \Msun$ &--&  $1.03_{-0.39}^{+0.47}$   &    $1.05_{-0.45}^{+0.53}$    \\ 
        M$_{200}$          & $M_{200; \ \rm MW}$   & $10^{11} \Msun$ &--&  $5.73_{-0.58}^{+0.76}$   &    $7.84_{-1.97}^{+3.08}$    \\ 
        R$_{200}$          & $R_{200; \ \rm MW}$   & $\kpc$          &--&  $170.7_{-5.9}^{+7.2 }$   &   $189.5_{-17.4}^{+22.1}$     \\  
        V$_\mathrm{escaped}$ at sun &$v_\mathrm{esc,\odot}$&$\kms$   &--&  $495.5_{-9.7}^{+11.2}$   &    $528.3_{-31.4}^{+55.3}$    \\
        DM density at sun  &$\rho_\mathrm{DM,\odot}$& GeV cm$^{-3}$  &--&  $0.34_{-0.01}^{+0.02}$   &    $0.34_{-0.02}^{+0.02}$     \\
        \hline\hline
        \\[-.3cm]
     \end{tabular}
\label{tab:FitPars}
\end{table*}

\subsection{Fits of the observational data}

Even though we have adopted the similar modeling method as in
\citet{Vasiliev2019a}, there are two major differences with them. First, we
have used Gaia EDR3 which improves the uncertainties by a factor of 2. Second,
we have added two important tight constraints by imposing the model to fit both the MW RC
and the vertical force data.  It would be useful to check how the DF different to
that of \citet{Vasiliev2019a}.

\begin{figure}
\center
\includegraphics[scale=0.53]{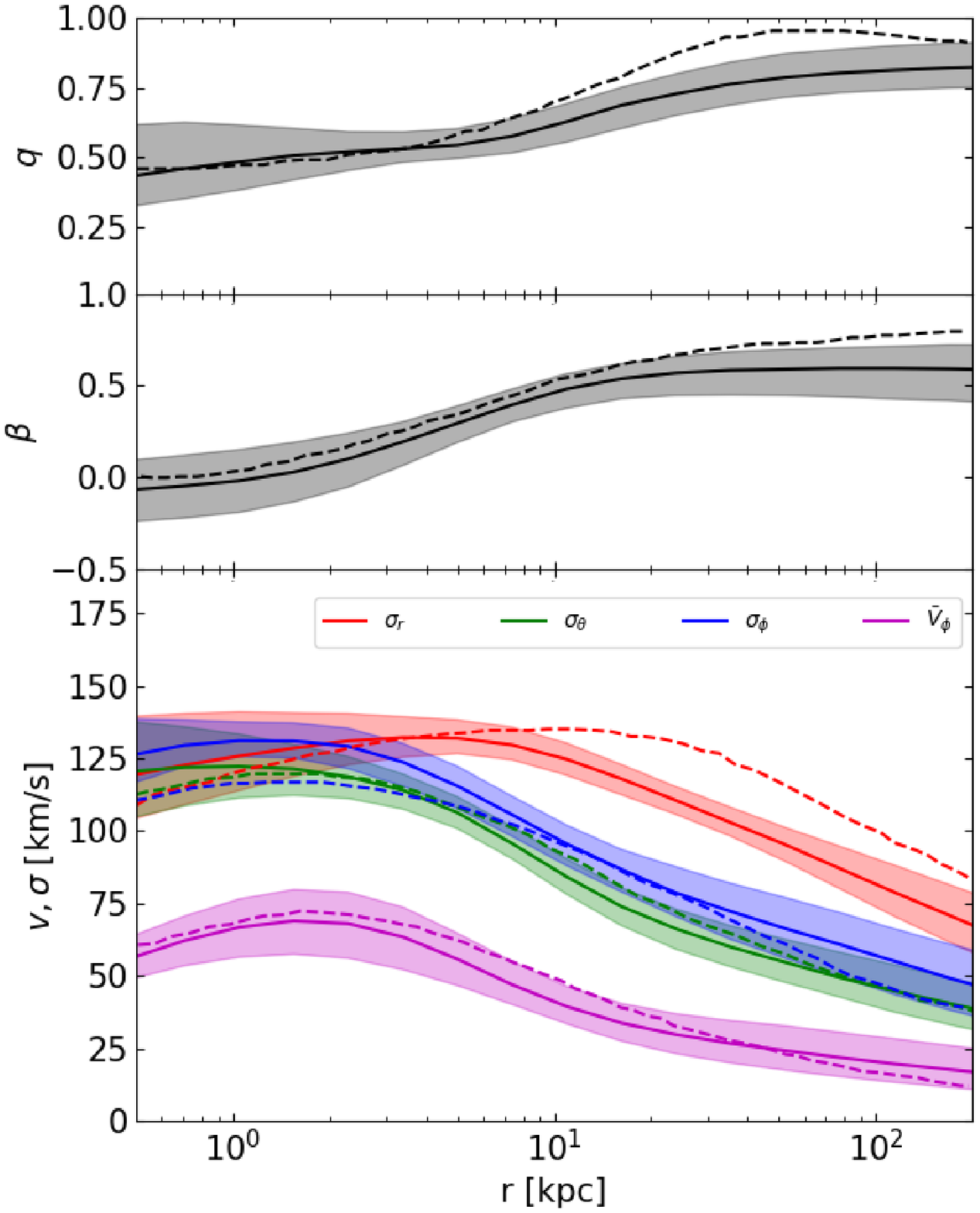}
\caption{Physical quantities estimated from ensemble of models from MCMC runs in function of radius. The solid lines are the mean
values estimated from MCMC models, while the shaded regions indicate the 68 per
cent credible regions.  Top panel: the axial ratio ($q=z/R$) of GCs spatial density profile varied as function
of radius.  Middle panel: the velocity
anisotropic parameter $\beta =
1-(\sigma^2_{\theta}+\sigma^2_{\phi})/(2\sigma^2_r)$ in function of the
radius.  Bottom panel: the velocity dispersions in three directions and the
mean azimuthal velocity as function of the radius. For comparison, the dotted-lines in each panel 
shows the results from \citet{Vasiliev2019b}. }
\label{fig:trVparam}
\end{figure}

Following \citet{Vasiliev2019a} we have derived from our posterior distribution
the velocity structure variation and the axis ratio of GCs as a function of the radius as
it is shown in Figure~\ref{fig:trVparam}. The velocity anisotropic parameter $\beta$
varies with radius, being isotropic in the center and radially dominated at the
outskirts.  The axis ratio $q$ increases with radius, which is consistent with
the disk component in the inner region \citep{Binney2017}.

Figure~\ref{fig:trVparam} shows our modeling results of GCs and compare them
to that of \citet{Vasiliev2019b}.  The velocity dispersion, anisotropic
parameter, and axis ratio of GCs are found to be very similar, however the radial velocity dispersion 
shows large discrepancy at $r> 10$ kpc from one to the other study.

\begin{figure}
\center
\includegraphics[scale=0.65]{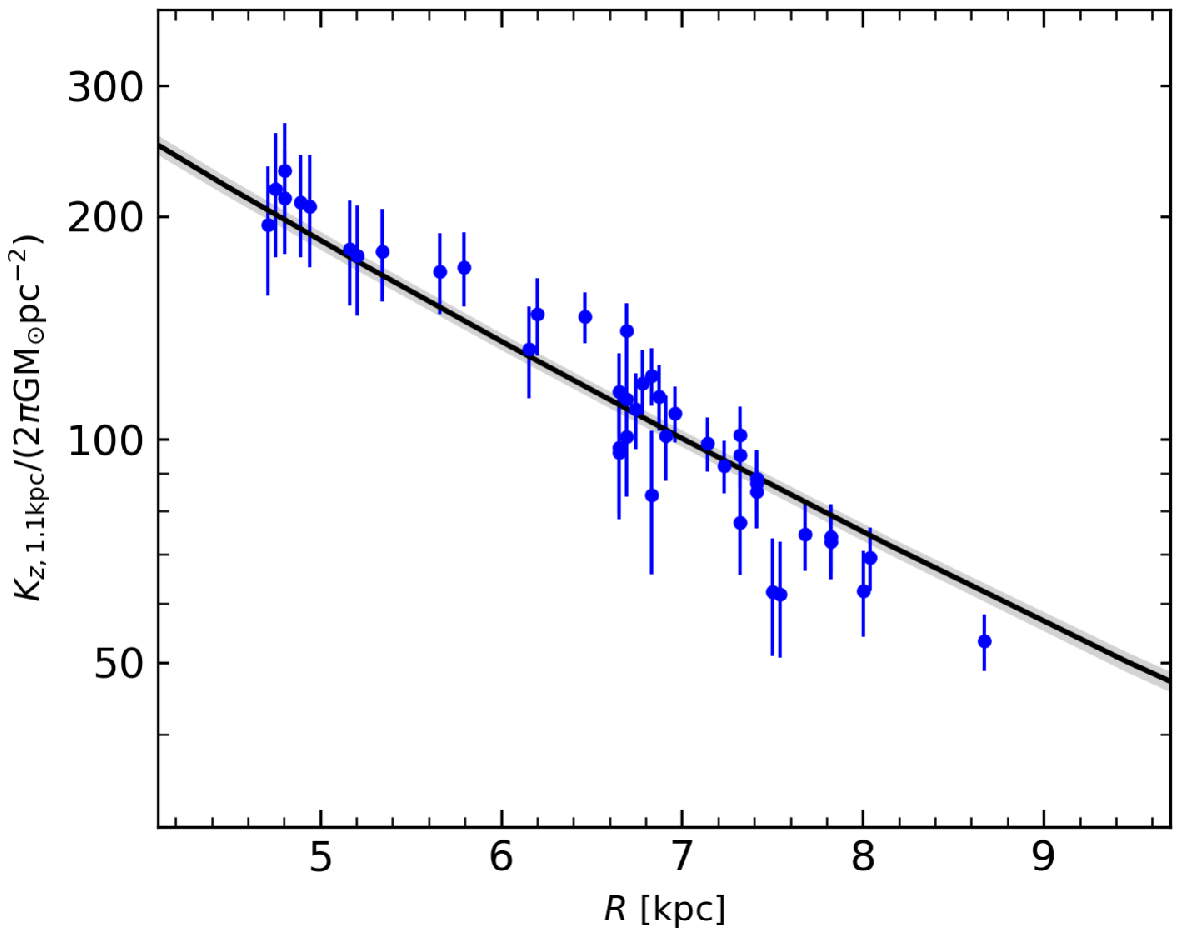}
\caption{The vertical force at $z=1.1$ kpc as function of radius. The blue
solid circle and error bars are from the observation measurements by
\citet{Bovy2013}, and the black line and shaded region indicate the estimated and 68
percentile of posterior distribution of our model.} 
\label{fig:Kz}
\end{figure}

Figure~\ref{fig:Kz} compares the vertical force $K_\mathrm{z=1.1 kpc}$ from 
the observed data \citep{Bovy2013} and that derived from our posterior distribution. 
The model reproduce well the observed vertical force.

\subsection{The Milky Way Rotation Curve}

\begin{figure*}
\center
\includegraphics[scale=1.2]{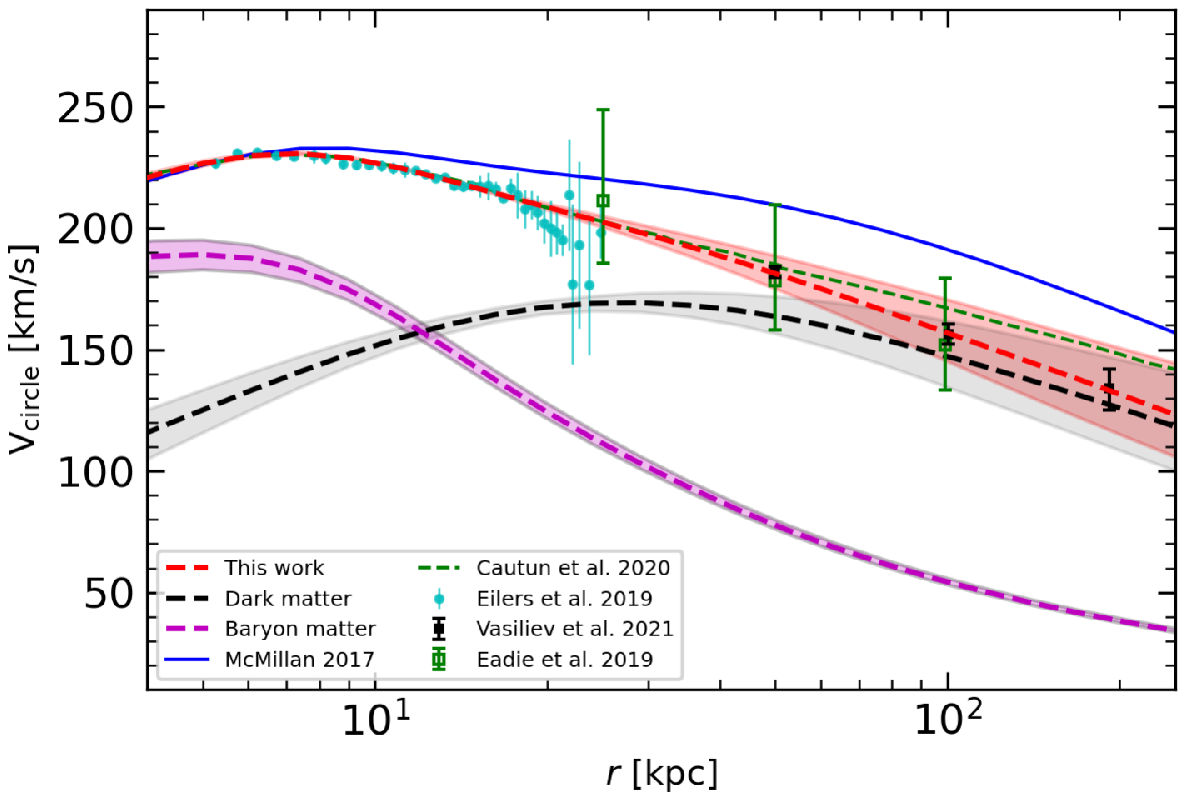}
\caption{It compares rotation curves derived from posterior distribution of our
models with literature. The Zhao's dark matter model is used. The shaded region indicate
the 68 percentile. The black dashed-line indicates the contribution from dark matter, while 
the magenta dashed-line shows the contribution from baryon matter. }
\label{fig:RC}
\end{figure*}

Figure~\ref{fig:RC} shows the RCs derived from our new modeling with Zhao's DM
density profiles (Eq.\ref{eq:Zhao}, the red-line and shaded region).  The
derived RC is fitting well the disk RC of \citet{Eilers2019}, for which velocities are
much lower than that predicted by \citet{McMillan2017}. The RC from Zhao's DM
profile is consistent with the recent results of \citet{Vasiliev2021a}
and \citet{Eadie2019}, as shown in Fig.\ref{fig:RC}.
\citet{Cautun2020} have also fitted the disk RC of \citet{Eilers2019} considering
the baryon contraction effect. They considered constraints from dwarf satellites
based on \citet{Callingham2019}, which results in a higher value of RC at outskirts than
our value.

\subsection{Escaped velocity and DM density at solar position}

\begin{figure}
\center
\includegraphics[scale=0.89]{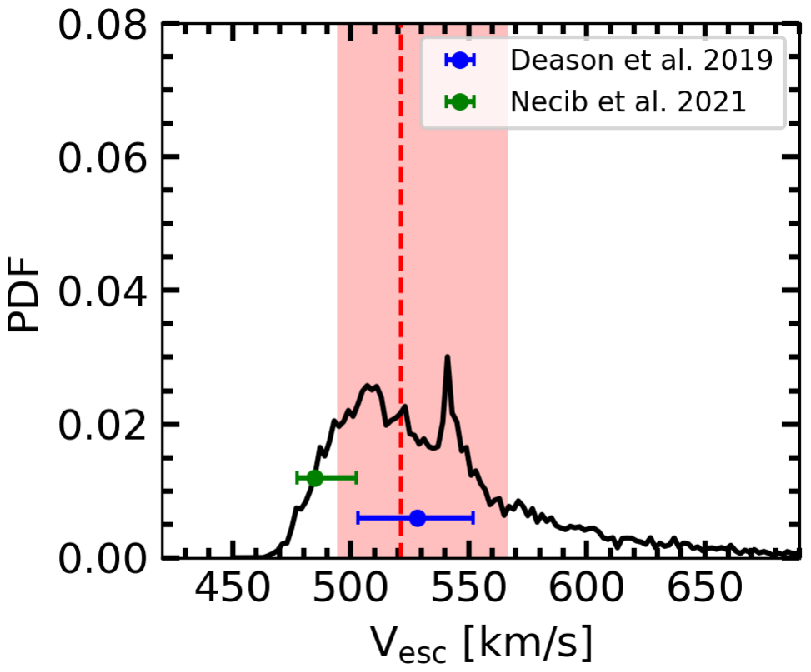}
\caption{The posterior distribution of the escaped velocity at the solar position
for the Zhao DM density model. The red-dashed line and shaded region indicate
the median and 68 percentiles for the distribution, and these values are listed
in Table \ref{tab:FitPars}} 
\label{fig:Vesc} 
\end{figure}

\begin{figure}
\center
\includegraphics[scale=0.59]{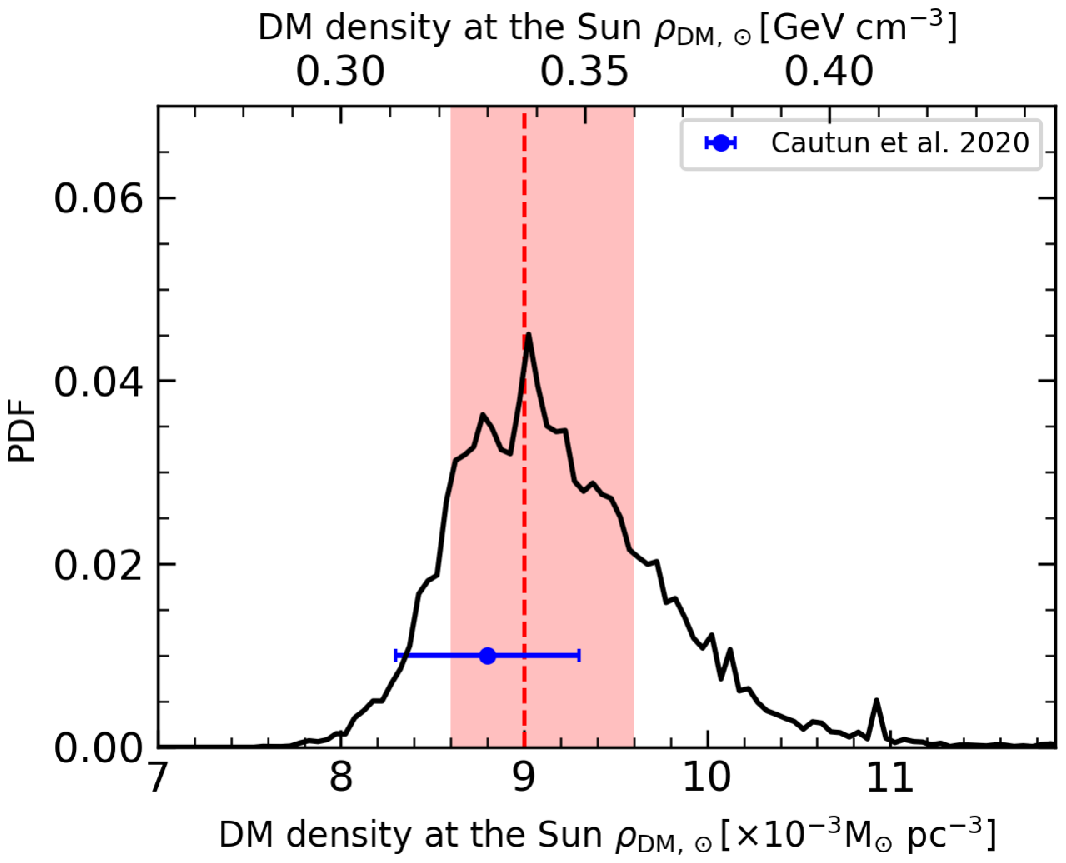}
\caption{The distribution of dark matter density at the solar position for the
Spheroid model. The median and associated 68 percentile of the posterior distribution from our MCMC runs are indicated by
red-dashed line and shaded regions, and these values are listed in Table
\ref{tab:FitPars}.} 
\label{fig:RhoDM} 
\end{figure}

Accurately deriving the DM mass density profile means that we can make predictions on
the escaped velocity and the DM density at solar position, and compare them with
different measurements in the literature. Figure~\ref{fig:Vesc} shows the probability distribution function (PDF) of the escaped velocity
at solar position. Our value is consistent with the most recent results on
the escaped velocity measurement \citep{Deason2019b,Necib2021}. Figure~\ref{fig:RhoDM} show the PDF of dark matter density at solar position. The
new results is consistent with results from \citet{Read2014}.

\section{Discussion}

\subsection{Influence of the a priori choice of the MW mass density profile}

The MW baryon content is relatively well known, though there are still
variations by $\sim$ 30\% for each component from one study to another (see
\citealt{Pouliasis2017}). Our method recovers these uncertainties by letting
varying by similar amount the baryonic components. \\

The DM content of the MW is less constrained since it is found highly dependent
on the choice of tracers. Here we have used very robust tracers, which are
stars embedded into the disk  and GCs, as well as constraints from the vertical
force. For the later, we even consider in the Appendix A the alternative for
which some (Crater) could be dwarf galaxy instead, or not bound (Pyxis). Our
goal is to keep as large as possible the range of DM profile for the halo. This
is why we have chosen both Zhao and Einasto profiles. The first one is a
generalization of the NFW and of the generalized NFW profiles that have been
often used to fit DM halos. The second is acknowledged to reproduce better the
DM halo density profile coming from simulations
\citep{Navarro2004,Dutton2014,Gao2008}. It has also the advantage to be
parameterized by only 3 parameters against 5 fo the Zhao profile, however it
may become a disadvantage if more complex DM distribution is required, for
e.g., fitting the MW mass profile in presence of a massive LMC. \\


\citet{Jiao2021} have shown that NFW and of the generalized NFW dark matter profiles may
be biased in favor of high values for the total MW mass when compared
to results using the Einasto profile. Results of our paper based on the Zhao's DM profile indeed provide higher mass values
than that from Einasto DM profile, which may confirm the \citet{Jiao2021} results.
Nevertheless we prefer to keep the whole range of
possibilities in fitting the DM component of the MW, and to consider the whole
range of MW masses provided by these two kinds of excellent models in
reproducing the DM.\\

A recent study shows that the DM profile could be changed during the process
of the baryon contraction in the center region, which result in profile deviate
from NFW \citep{Cautun2020}. We do not think this can alter our conclusions,
because \citet{Jiao2021} showed that the Einasto model is able to
reproduce a contracting halo.



\subsection{A massive LMC may introduce disequilibrium}

There has been many clues that a massive LMC $\sim 10^{11}\Msun$ passing
by MW could have non-negligible effects on the MW 
\citep{Erkal2020,Petersen2020,Conroy2021}, and on the track 
 of stellar streams \citep{Vasiliev2021a,Erkal2019, Koposov2019}. For example,
several halo tracers have shown velocity gradients that are predicted by a massive
LMC model \citep{Petersen2020}.

However, the fact that the LMC could very massive is still under discussion.
For example,  GC distributions show no significant velocity shift
\citep{Erkal2020}. \citet{Conroy2021} found that halo K giants show a local
wake and a Northern over-density, which can be explained by the passage of a
massive LMC. However, as they showed a reasonable tilted triaxial halo model
can explain this phenomenon equally well. It has often acknowledged that the LMC
is at first-passage to the MW \citep{Kallivayalil2013}. The splendid Magellanic
Stream has been well reproduced under the frame of 'ram-pressure plus
collision' model \citep{Hammer2015,Wang2019}, which reproduces well the neutral
gas morphology including its structure into two filaments, the observed hot
ionized distribution, as well as the very peculiar stellar morphology of the
SMC. This model requires the total mass of LMC to be less than $2\times
10^{10}$ M$_{\odot}$, which is almost one decade smaller than that of a very
massive LMC.  

To test the effect of LMC on the final MW mass measurement, we make a
simulation to test how a massive LMC passing by MW may affect mass estimates.
We follow the same method of \citet{Vasiliev2021a} and we have built the pair
of MW and LMC. \citet{Vasiliev2021a} built models of the MW and LMC interaction
to investigate the effect on the Sagittarius stream track. Their MW model
consist of a stellar disk, bulge, and dark matter halo. The dark matter halo
have axis-symmetric or triaxial-symmetric shape. For simplicity, we have used a
spherical dark matter model. We also introduce a light gas component, which
does not produce an essential effect on the total mass profile, but that is
used to generate test particles for reproducing the modeled rotation curve.  The LMC
has a truncated NFW profile with total mass 1.5$\times 10^{11}$ M$_{\odot}$.
We notice that the model of Vasiliev et al (2021) does not reproduce the MW RC
(Eilers et al. 2019) and that it overestimates rotational velocities by about 5\%. We then slightly
scale down the MW mass value to match the rotation curve.  We note that these
small change have little effects on the final results.  We also remark that
this modeling does not intent to reproduce whole full properties of the MW and
massive LMC. Instead, its goal is to gauge the effect of a massive LMC to the
constraints from the MW RC.  Details on the structure of the pairs of LMC and
MW and on the simulations of their interactions can be found in
\citet{Vasiliev2021a}.

\begin{figure*}
\center
\includegraphics[scale=0.45]{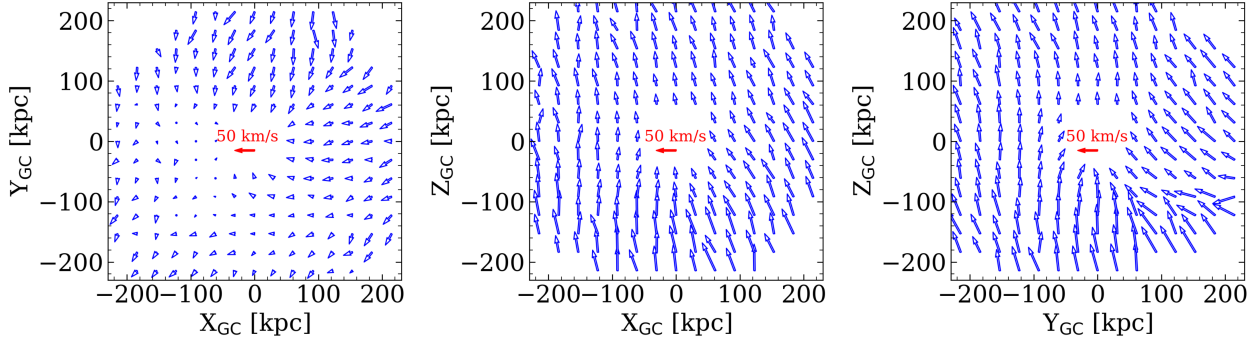}
\caption{The simulated velocity vectors maps for dark matter halo particles at different positions after 
the perturbation of a massive LMC (1.5$\times 10^{11}$ M$_{\odot}$) passing by. Only particles outside the
disk region (r$>$20 kpc) are shown. }
\label{fig:Vmap}
\end{figure*}

The LMC starts from 427 kpc away and is launched to reach the current observed
position, at about 50 kpc to the MW center. The top row of
Figure~\ref{fig:Vmap} shows the final velocity vector map of MW dark matter
particles. The massive LMC induces a strong disequilibrium for MW system in
which the systematic velocities are changed at different positions, which
indicates that correcting the systematic effect is complex. 


Bearing in mind that the GC system shows no systematic velocity shift
\citep{Erkal2020}, we build a mock observational sample from the simulated  
samples (Figure~\ref{fig:Vmap}). We randomly select the mock GC sample 
by using dark matter halo particles, adjusting their number to that of GCs.

With the simulated (mock) GC samples, we have perturbed the true value ($s,
v_{\los}, \mu_\alpha^*, \mu_\delta$) according to the uncertainties of the
observed GCs.  The distance uncertainty is $\sim 5\%$.  The mean errors of
line-of-sight velocity of the GCs are very small and fixed to 1.8 km/s.  The
mean proper motion errors in both directions are $\sim 0.03$ mas/yr. The
covariance correlation coefficiency are set by a randomly selected from
Gaussian distribution with $\sigma = 0.06$ and zero mean value following observations.  

To mimic the observed rotation curve of \citet{Eilers2019}, we have used the
mean streaming velocity of gas particle, which has less velocity dispersion,
and then less asymmetric drift correction. We also measured the vertical force
at 1.1 kpc above the disk, which mimics the observation data \citep{Bovy2013}.
The RC and vertical force data in the simulated model have similar fraction
errors as observations.

\begin{figure*}
\center
\includegraphics[scale=1.2]{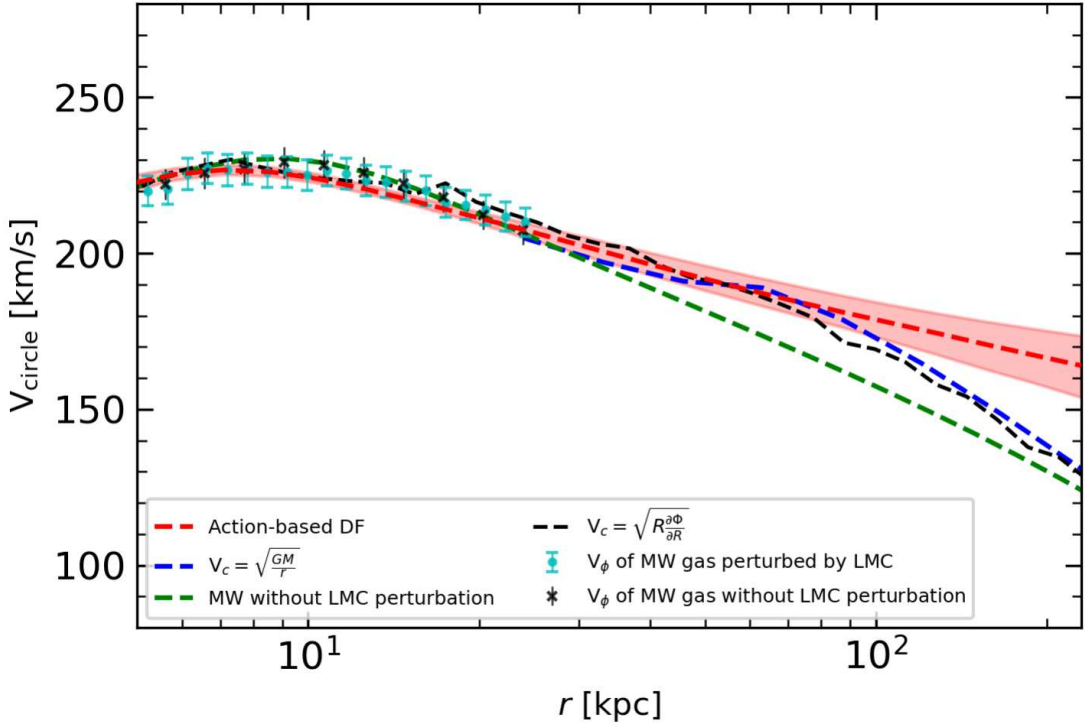}
\caption{It compares the measured RC for the simulated GCs with observed values. The
green-dashed line show the input RC of MW without LMC perturbation. The
blue-dahsed line shows the rotation curve for the MW perturbed by LMC
calculated with V$_c=\sqrt{\frac{GM(<r)}{r}}$, which include the contribution
by LMC. The black crosses show the measured streaming velocity of gas without
LMC perturbation. The cyan-dots show the streaming velocity of gas disk
after LMC perturbation. The red-dash line shows the RC recovered by the
action-based distribution method with shaded region indicate 68 per cent 
credible regions. The black-dashed line indicates the RC derived with
${\sqrt{R\frac{\partial{\Phi}}{\partial{R}}}}$ after LMC perturbation. }
\label{fig:simRC}
\end{figure*}

With the mimicked observation data in hand, we have used the action-based DF
method listed above to model the gravitational field on the simulated data.
Figure~\ref{fig:simRC} compares the final result to the true values. The green
dashed-line shows the true rotation curve of input MW without LMC perturbation.
The blue dashed-line indicates the rotation curve derived with
V$_c=\sqrt{\frac{GM(<r)}{r}}$ assuming a spherical mass distribution for the
overall contribution of MW and LMC. The black-dashed line indicates the RC
derived with $\sqrt{R\frac{\partial{\Phi}}{\partial{R}}}$.  The mass estimate
with the spherical mass distribution assumption is less accurate, because of the
non-spherical shape of disk mass distribution and of the  LMC contribution. The
red-dashed line shows the results with action-based DF modeling. The
contribution from the massive LMC is well recovered by the action-based PDF
modeling, as shown by the slight bump of RC at $\sim 50$ kpc in Figure~\ref{fig:simRC}. The
introduction of a massive LMC leads to overestimate the mass at large radius
($r>100$ kpc).  The black and cyan symbols show the gas streaming velocity in
the center region for both without and with LMC perturbation, both of these
rotation velocities are consistent with each other. This indicates that the
central region within the disk is much less affected by the massive LMC than
the outer halo region, which provides us further confidence in using of
rotation curve data from \citet{Eilers2019}.



\subsection{Effect of substructures} \label{sec:unrelaxed}

A recent discovery shows that the MW halo consists of many substructures, for
example, the Sagittarius streams which contribute large fraction of halo stars
\citep[$10\sim 15$ per cent;][]{Deason2021,Deason2019a}. The big merger event,
$Gaia$ Sausage or $Gaia$ Enceladus, which occurred 10 Gyr ago also contributes
to a large fraction of inner halo stars
\citep{Belokurov2018,Helmi2018,Naidu2021}

\begin{figure}
\center
\includegraphics[scale=0.73]{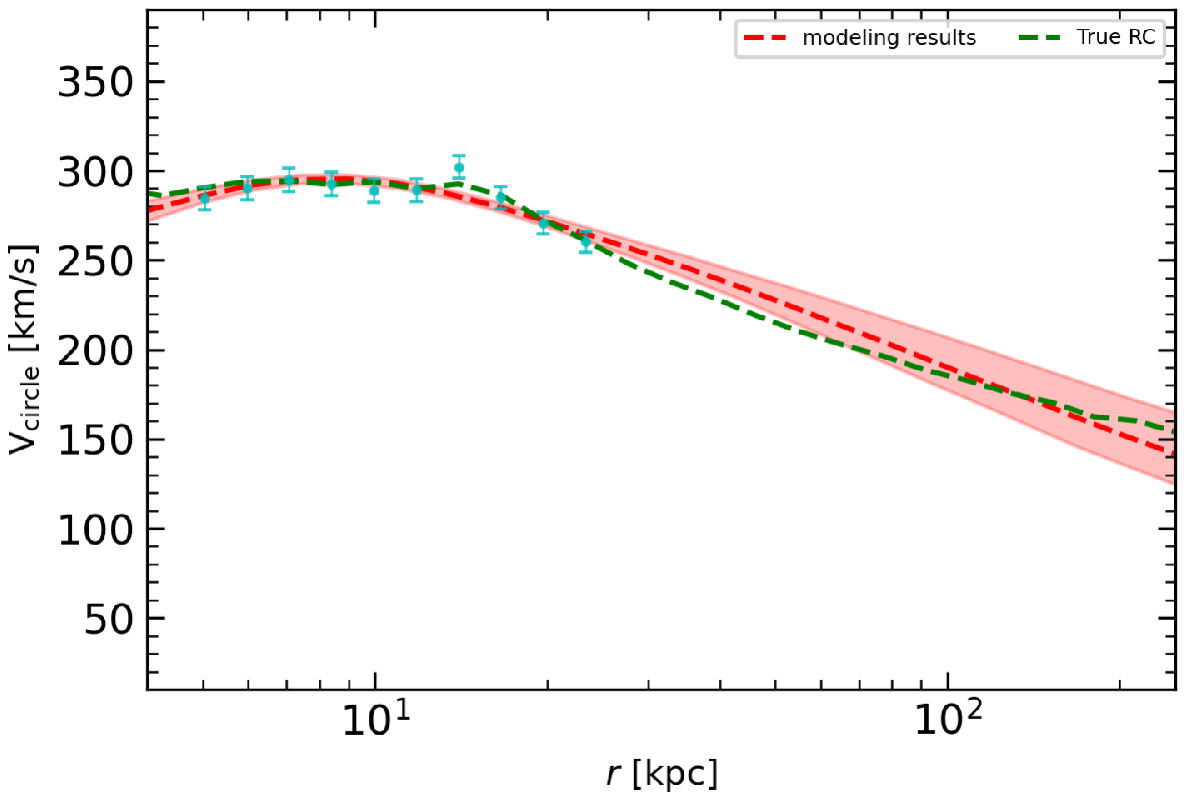}
\caption{Comparing modeling results with mock MW-like of model 'm12m' of FIRE2, which 
is a MW-like galaxy from cosmological hydrodynamic simulations with unrelaxed substructures. 
The green-dashed line indicates the true rotation curve for the model 'm12m', while 
the cyan points show the rotation curve data with random errors added following observational 
errors. The red-dashed line and pink shaded region denote the modeling results and 68 percent credible region 
from the models of MCMC run.} 
\label{fig:m12m} 
\end{figure}

In order to test the effect of these unrelaxed substructure effect on the
measurement, we use the model m12m of the FIRE2 Latte cosmological hydrodynamic
simulations suite, which produces a realistic Milky-Way-like galaxy, including
many unrelaxed substructures \citep{Wetzel2016,Hopkins2015,Hopkins2018}. From
this model, we generate the mock GC sample.  We select stars from model
'm12m' with age older than 10 Gyr, using them to represent the GC samples. From
the simulated model we derive the rotation curve and vertical force at 1.1 kpc
and add observational errors as in \citet{Hattori2021}.  With our modeling
machine we derive the final rotation curve and compare it with input data as
shown in Fig.  \ref{fig:m12m}.  The unrelaxed substructures in the halo result
only in moderate fluctuations of the rotation curve. 
 
\subsection{MW total mass and comparison with literature and implication for cosmology}

The total mass is critical for many cosmological satellite problem, for
instance, "too-big-to-fail" \citep{Boylan-Kolchin2011, Wang2012}, and missing
satellite problem.  Our measurement for the mass of the MW is
$7.84_{-1.97}^{+3.08} \times 10^{11}$ M$\odot$ and
$5.8_{-0.68}^{+0.81}\times10^{11}$ M$_{\odot}$, after using the Zhao and the Einasto
model for DM, respectively. Appendix discusses how these values can be slightly
affected by different ways in using the GC sample, i.e., by removing or not
Crater and Pyxis.


\begin{figure}
\center
\includegraphics[scale=0.92]{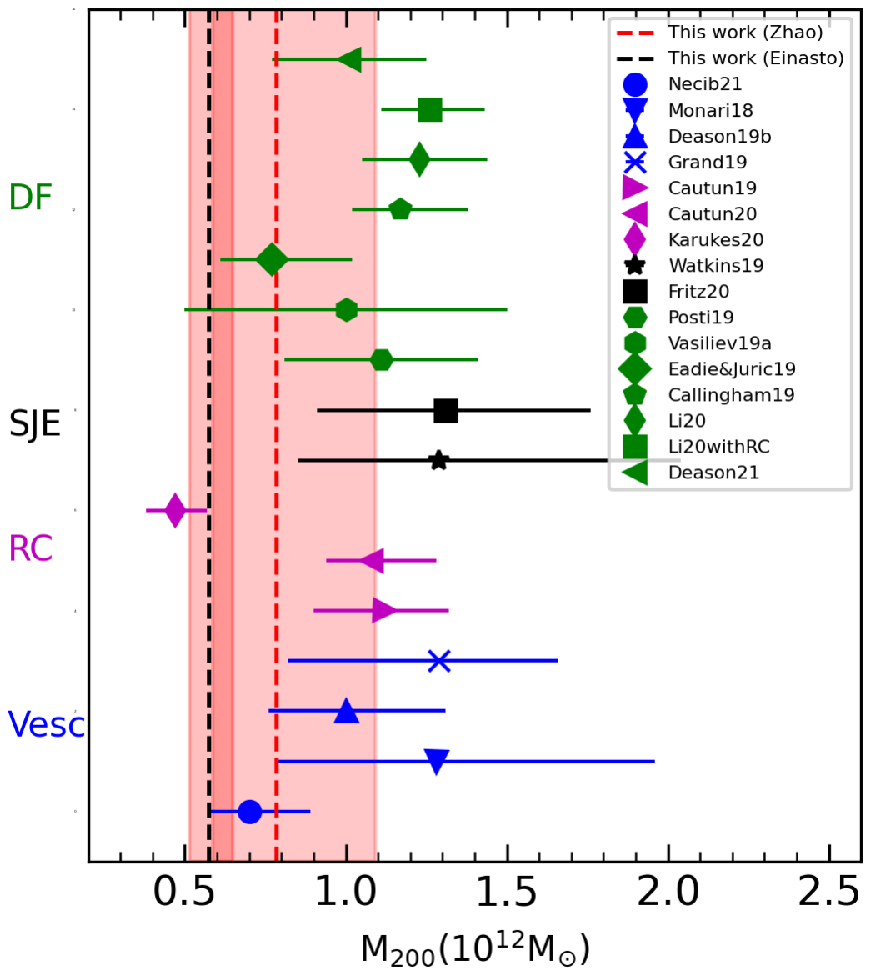}
\caption{Comparing MW total mass results measured in this work with that made use of $Gaia$ DR2 or EDR3. 
This figure is an update of Fig.5 in \citet{Wang2020}. Different methods 
has been labeled with different color. DF(distribution function):
\citet{Posti2019,Vasiliev2019a,Eadie2019,Callingham2019,Li2020, Deason2021},
Spherical Jeans Equation (SJE): \citet{Watkins2019,Fritz2020}, RC:
\citet{Karukes2020,Cautun2019, Cautun2020}, Escaped Velocity(Vesc):
\citet{Necib2021,Monari2018,Deason2019b,Grand2019}. Red-dashed line indicates
result of this work for Zhao's DM profile, and black-dashed line show result from Einasto's DM profile. 
The pink shaded region shows the 68 percentile credible intervals.} 
\label{fig:Mtot} 
\end{figure}

Figure~\ref{fig:Mtot} compares the total MW mass measured in this work with
recent results by using $Gaia$ DR2 and EDR3.  This figure is an update of
Fig. 5 of \citet{Wang2020}, in which the results are grouped on the basis of the different
methods used to estimate the MW total mass. Our range of estimates is at the
low end of MW mass, which may alleviate the tension of the 'too-big-to fail'
problem. Recent studies have suggested that only three MW satellites (MCs and
Sagittarius dwarf, see \citealt{Wang2012}) could inhabit in sub-halos with their value
of V$_\mathrm{max}$ larger than a threshold V$_{\mathrm{th}} \sim 30$ km/s ,
which is defined by \citet{Boylan-Kolchin2011} as the massive failure
threshold. \citet{Wang2012} used $\Lambda$CDM cosmological simulations
and showed that only $\sim 5$ per cent of halos with mass $M_{\mathrm{halo}}\sim
2\times 10^{12}$ M$_{\odot}$ have three or fewer sub-haloes with
V$_{\mathrm{max}} > 30$ km/s, while this fraction increases to $\sim 70$ per
cent for a MW mass of $7.5\times10^{11}$ M$_{\odot}$. The total mass of MW
in our measurement naturally includes the contribution for LMC in our
measurement, since the contribution by LMC has been added into the measured
velocity for GCs.  By assuming the LMC mass is 1.5$\times 10^{11}$ M$_{\odot}$,
leads to a MW total mass of 6.34$\times 10^{11}$ M$_{\odot}$. 



\section{CONCLUSION}

Using $Gaia$ EDR3 data, we derive proper motions for about 150 MW GCs. When
comparing their proper motions with that from $Gaia$ DR2, 
errors decrease by about a factor 2, which is consistent with the $Gaia$ data
reduction analysis. 

With the newly derived proper motions for the MW GCs and by combining them to the
constraints from the rotation curve from 5 to 25 kpc and from the vertical
force measurements, we have built dynamical models for the MW using the
action-based distribution function. From the new dynamical model we have
derived the rotation curve and the mass profile for MW, and have compared them with
recent results based on Gaia data.  The local dark matter density and local
escaped velocity are all consistent with literature values. 

We have used mock simulation data to test the robustness of our results.
Firstly, we consider the perturbation of a possible massive LMC passing by MW,
which results in the reflex motion of halo stars with velocity intensities and
directions modified at different positions (Figure~\ref{fig:Vmap}). By modeling
mock GCs system from the simulations with action-based DF and comparing with
the input value, we found the modeling can well recover the input rotation
curve value including the contribution from the massive LMC ($1.5\times
10^{11}$ M$_{\odot}$) within 100 kpc. At large distances, this model
overestimates the rotation curve $\sim 20$ percent at 200 kpc.  Secondly, we
consider the effect of unrelaxed substructures on the results. We have used the
realistic cosmological hydrodynamic simulations from FIRE2 Latter simulation
data suite. The model "m12m" produce a MW-like galaxy with unrelaxed
substructures. From the data model, we select stars with age older than 10 Gyrs
to build mock GCs system. The unrelaxed substructure results in the final
rotation curve fluctuating around its true value by about 10 percent.\\

In this paper we have chosen the most objective view in adopting baryonic and
DM mass, by avoiding a priori against or for a given modeling. It results that
the total mass of the MW ranges from $5.36_{-0.68}^{+0.81}\times10^{11}$
M$_{\odot}$ to $7.84_{-1.97}^{+3.08} \times 10^{11}$ M$\odot$, which
significantly narrows the previous ranges for the MW mass in the literature. 


\section*{Acknowledgments}

We thanks the referee helpful comments, which have significantly improved the manuscript.
The computing task was carried out on the HPC cluster at China National
Astronomical Data Center (NADC). NADC is a National Science and Technology
Innovation Base hosted at National Astronomical Observatories, Chinese Academy
of Sciences. This work is supported by Grant No. 12073047 of the National
Natural Science Foundation of China.

This work has made use of data from the European Space Agency (ESA)
mission {\it Gaia}
(\url{http://www.cosmos.esa.int/gaia}),
processed by the {\it Gaia} Data Processing and Analysis Consortium (DPAC,
\url{http://www.cosmos.esa.int/web/gaia/dpac/consortium}).
Funding for the DPAC has been provided by national institutions, in particular
the institutions participating in the {\it Gaia} Multilateral Agreement.

\section*{DATA AVAILABILITY}

The data underlying this article will be shared on reasonable request
to the corresponding author.

\bibliographystyle{mn2e}
\bibliography{reference.bib}

\providecommand{\noopsort}[1]{}\providecommand{\singleletter}[1]{#1}%
\begin{thebibliography}{97}
\expandafter\ifx\csname natexlab\endcsname\relax\def\natexlab#1{#1}\fi

\bibitem[{{Ablimit} {et~al}\mbox{.}(2020){Ablimit}, {Zhao}, {Flynn}, \&
  {Bird}}]{Ablimit2020}
{Ablimit} I., {Zhao} G., {Flynn} C., {Bird} S.~A., 2020, ApJL, 895, L12

\bibitem[{{Belokurov} {et~al}\mbox{.}(2018){Belokurov}, {Erkal}, {Evans},
  {Koposov}, \& {Deason}}]{Belokurov2018}
{Belokurov} V., {Erkal} D., {Evans} N.~W., {Koposov} S.~E., {Deason} A.~J.,
  2018, MNRAS, 478, 611

\bibitem[{{Binney}(2020)}]{Binney2020}
{Binney} J., 2020, in Galactic Dynamics in the Era of Large Surveys, {Valluri}
  M., {Sellwood} J.~A., eds., Vol. 353, pp. 101--108

\bibitem[{{Binney} \& {Wong}(2017)}]{Binney2017}
{Binney} J., {Wong} L.~K., 2017, MNRAS, 467, 2446

\bibitem[{{Bland-Hawthorn} \& {Gerhard}(2016)}]{Bland-Hawthorn2016}
{Bland-Hawthorn} J., {Gerhard} O., 2016, ARA\&A, 54, 529

\bibitem[{{Bonifacio} {et~al}\mbox{.}(2015){Bonifacio}, {Caffau}, {Zaggia},
  {Fran{\c{c}}ois}, {Sbordone}, {Andrievsky}, \& {Korotin}}]{Bonifacio2015}
{Bonifacio} P., {Caffau} E., {Zaggia} S., {Fran{\c{c}}ois} P., {Sbordone} L.,
  {Andrievsky} S.~M., {Korotin} S.~A., 2015, A\&A, 579, L6

\bibitem[{{Bovy}(2015)}]{Bovy2015}
{Bovy} J., 2015, ApJS, 216, 29

\bibitem[{{Bovy} {et~al}\mbox{.}(2016){Bovy}, {Bahmanyar}, {Fritz}, \&
  {Kallivayalil}}]{Bovy2016}
{Bovy} J., {Bahmanyar} A., {Fritz} T.~K., {Kallivayalil} N., 2016, ApJ, 833, 31

\bibitem[{{Bovy} \& {Rix}(2013)}]{Bovy2013}
{Bovy} J., {Rix} H.-W., 2013, ApJ, 779, 115

\bibitem[{{Bowden}, {Belokurov} \& {Evans}(2015){Bowden}, {Belokurov}, \&
  {Evans}}]{Bowden2015}
{Bowden} A., {Belokurov} V., {Evans} N.~W., 2015, MNRAS, 449, 1391

\bibitem[{{Boylan-Kolchin}, {Bullock} \& {Kaplinghat}(2011){Boylan-Kolchin},
  {Bullock}, \& {Kaplinghat}}]{Boylan-Kolchin2011}
{Boylan-Kolchin} M., {Bullock} J.~S., {Kaplinghat} M., 2011, MNRAS, 415, L40

\bibitem[{{Bullock} \& {Boylan-Kolchin}(2017)}]{Bullock2017}
{Bullock} J.~S., {Boylan-Kolchin} M., 2017, ARA\&A, 55, 343

\bibitem[{{Callingham} {et~al}\mbox{.}(2019){Callingham}, {Cautun}, {Deason},
  {Frenk}, {Wang}, {G{\'o}mez}, {Grand}, {Marinacci}, \&
  {Pakmor}}]{Callingham2019}
{Callingham} T.~M. {et~al.}, 2019, MNRAS, 484, 5453

\bibitem[{{Cautun} {et~al}\mbox{.}(2020){Cautun}, {Ben{\'\i}tez-Llambay},
  {Deason}, {Frenk}, {Fattahi}, {G{\'o}mez}, {Grand}, {Oman}, {Navarro}, \&
  {Simpson}}]{Cautun2020}
{Cautun} M. {et~al.}, 2020, MNRAS, 494, 4291

\bibitem[{{Cautun} {et~al}\mbox{.}(2019){Cautun}, {Deason}, {Frenk}, \&
  {McAlpine}}]{Cautun2019}
{Cautun} M., {Deason} A.~J., {Frenk} C.~S., {McAlpine} S., 2019, MNRAS, 483,
  2185

\bibitem[{{Cautun} {et~al}\mbox{.}(2014){Cautun}, {Hellwing}, {van de
  Weygaert}, {Frenk}, {Jones}, \& {Sawala}}]{Cautun2014}
{Cautun} M., {Hellwing} W.~A., {van de Weygaert} R., {Frenk} C.~S., {Jones} B.
  J.~T., {Sawala} T., 2014, MNRAS, 445, 1820

\bibitem[{{Chua} {et~al}\mbox{.}(2019){Chua}, {Pillepich}, {Vogelsberger}, \&
  {Hernquist}}]{Chua2019}
{Chua} K. T.~E., {Pillepich} A., {Vogelsberger} M., {Hernquist} L., 2019,
  MNRAS, 484, 476

\bibitem[{{Conroy} {et~al}\mbox{.}(2021){Conroy}, {Naidu}, {Garavito-Camargo},
  {Besla}, {Zaritsky}, {Bonaca}, \& {Johnson}}]{Conroy2021}
{Conroy} C., {Naidu} R.~P., {Garavito-Camargo} N., {Besla} G., {Zaritsky} D.,
  {Bonaca} A., {Johnson} B.~D., 2021, Nature, 592, 534

\bibitem[{{Cui} {et~al}\mbox{.}(2012){Cui}, {Zhao}, {Chu}, {Li}, {Li}, {Zhang},
  {Su}, {Yao}, {Wang}, {Xing}, {Li}, {Zhu}, {Wang}, {Gu}, {Luo}, {Xu}, {Zhang},
  {Liu}, {Zhang}, {Yang}, {Cao}, {Chen}, {Chen}, {Chen}, {Chen}, {Chu}, {Feng},
  {Gong}, {Hou}, {Hu}, {Hu}, {Hu}, {Jia}, {Jiang}, {Jiang}, {Jiang}, {Jin},
  {Li}, {Li}, {Li}, {Liu}, {Liu}, {Lu}, {Mao}, {Men}, {Qi}, {Qi}, {Shi},
  {Tang}, {Tao}, {Wang}, {Wang}, {Wang}, {Wang}, {Wang}, {Wang}, {Wang},
  {Wang}, {Wang}, {Wang}, {Wang}, {Wang}, {Xu}, {Xu}, {Yang}, {Yu}, {Yuan},
  {Yuan}, {Zhai}, {Zhang}, {Zhang}, {Zhang}, {Zhao}, {Zhou}, {Zhou}, {Zhu}, \&
  {Zou}}]{Cui2012}
{Cui} X.-Q. {et~al.}, 2012, Research in Astronomy and Astrophysics, 12, 1197

\bibitem[{{Das} \& {Binney}(2016)}]{Das2016a}
{Das} P., {Binney} J., 2016, MNRAS, 460, 1725

\bibitem[{{Das}, {Williams} \& {Binney}(2016){Das}, {Williams}, \&
  {Binney}}]{Das2016b}
{Das} P., {Williams} A., {Binney} J., 2016, MNRAS, 463, 3169

\bibitem[{{Deason}, {Belokurov} \& {Sanders}(2019){Deason}, {Belokurov}, \&
  {Sanders}}]{Deason2019a}
{Deason} A.~J., {Belokurov} V., {Sanders} J.~L., 2019, MNRAS, 490, 3426

\bibitem[{{Deason} {et~al}\mbox{.}(2021){Deason}, {Erkal}, {Belokurov},
  {Fattahi}, {G{\'o}mez}, {Grand}, {Pakmor}, {Xue}, {Liu}, {Yang}, {Zhang}, \&
  {Zhao}}]{Deason2021}
{Deason} A.~J. {et~al.}, 2021, MNRAS, 501, 5964

\bibitem[{{Deason} {et~al}\mbox{.}(2019){Deason}, {Fattahi}, {Belokurov},
  {Evans}, {Grand}, {Marinacci}, \& {Pakmor}}]{Deason2019b}
{Deason} A.~J., {Fattahi} A., {Belokurov} V., {Evans} N.~W., {Grand} R. J.~J.,
  {Marinacci} F., {Pakmor} R., 2019, MNRAS, 485, 3514

\bibitem[{{Di Matteo} {et~al}\mbox{.}(2008){Di Matteo}, {Bournaud}, {Martig},
  {Combes}, {Melchior}, \& {Semelin}}]{Matteo2008}
{Di Matteo} P., {Bournaud} F., {Martig} M., {Combes} F., {Melchior} A.~L.,
  {Semelin} B., 2008, A\&A, 492, 31

\bibitem[{{Dutton} \& {Macci{\`o}}(2014)}]{Dutton2014}
{Dutton} A.~A., {Macci{\`o}} A.~V., 2014, MNRAS, 441, 3359

\bibitem[{{Eadie} \& {Juri{\'c}}(2019)}]{Eadie2019}
{Eadie} G., {Juri{\'c}} M., 2019, ApJ, 875, 159

\bibitem[{{Eilers} {et~al}\mbox{.}(2019){Eilers}, {Hogg}, {Rix}, \&
  {Ness}}]{Eilers2019}
{Eilers} A.-C., {Hogg} D.~W., {Rix} H.-W., {Ness} M.~K., 2019, ApJ, 871, 120

\bibitem[{{Einasto}(1965)}]{Einasto1965}
{Einasto} J., 1965, Trudy Astrofizicheskogo Instituta Alma-Ata, 5, 87

\bibitem[{{Erkal} {et~al}\mbox{.}(2019){Erkal}, {Belokurov}, {Laporte},
  {Koposov}, {Li}, {Grillmair}, {Kallivayalil}, {Price-Whelan}, {Evans},
  {Hawkins}, {Hendel}, {Mateu}, {Navarro}, {del Pino}, {Slater}, {Sohn}, \&
  {Orphan Aspen Treasury Collaboration}}]{Erkal2019}
{Erkal} D. {et~al.}, 2019, MNRAS, 487, 2685

\bibitem[{{Erkal}, {Belokurov} \& {Parkin}(2020){Erkal}, {Belokurov}, \&
  {Parkin}}]{Erkal2020}
{Erkal} D., {Belokurov} V.~A., {Parkin} D.~L., 2020, MNRAS, 498, 5574

\bibitem[{{Fabricius} {et~al}\mbox{.}(2021){Fabricius}, {Luri}, {Arenou},
  {Babusiaux}, {Helmi}, {Muraveva}, {Reyl{\'e}}, {Spoto}, {Vallenari},
  {Antoja}, {Balbinot}, {Barache}, {Bauchet}, {Bragaglia}, {Busonero},
  {Cantat-Gaudin}, {Carrasco}, {Diakit{\'e}}, {Fabrizio}, {Figueras},
  {Garcia-Gutierrez}, {Garofalo}, {Jordi}, {Kervella}, {Khanna}, {Leclerc},
  {Licata}, {Lambert}, {Marrese}, {Masip}, {Ramos}, {Robichon}, {Robin},
  {Romero-G{\'o}mez}, {Rubele}, \& {Weiler}}]{Fabricius2020}
{Fabricius} C. {et~al.}, 2021, A\&A, 649, A5

\bibitem[{{Foreman-Mackey} {et~al}\mbox{.}(2013){Foreman-Mackey}, {Hogg},
  {Lang}, \& {Goodman}}]{Foreman-Mackey2013}
{Foreman-Mackey} D., {Hogg} D.~W., {Lang} D., {Goodman} J., 2013, PASP, 125,
  306

\bibitem[{{Fritz} {et~al}\mbox{.}(2020){Fritz}, {Di Cintio}, {Battaglia},
  {Brook}, \& {Taibi}}]{Fritz2020}
{Fritz} T.~K., {Di Cintio} A., {Battaglia} G., {Brook} C., {Taibi} S., 2020,
  MNRAS, 494, 5178

\bibitem[{{Fritz} {et~al}\mbox{.}(2017){Fritz}, {Linden}, {Zivick},
  {Kallivayalil}, {Beaton}, {Bovy}, {Sales}, {Sohn}, {Angell},
  {Boylan-Kolchin}, {Carrasco}, {Damke}, {Davies}, {Majewski}, {Neichel}, \&
  {van der Marel}}]{Fritz2017}
{Fritz} T.~K. {et~al.}, 2017, ApJ, 840, 30

\bibitem[{{Gaia Collaboration} {et~al}\mbox{.}(2020){Gaia Collaboration},
  {Brown}, {Vallenari}, {Prusti}, {de Bruijne}, {Babusiaux}, \&
  {Biermann}}]{Brown2020}
{Gaia Collaboration}, {Brown} A.~G.~A., {Vallenari} A., {Prusti} T., {de
  Bruijne} J.~H.~J., {Babusiaux} C., {Biermann} M., 2020, arXiv e-prints,
  arXiv:2012.01533

\bibitem[{{Gao} {et~al}\mbox{.}(2008){Gao}, {Navarro}, {Cole}, {Frenk},
  {White}, {Springel}, {Jenkins}, \& {Neto}}]{Gao2008}
{Gao} L., {Navarro} J.~F., {Cole} S., {Frenk} C.~S., {White} S. D.~M.,
  {Springel} V., {Jenkins} A., {Neto} A.~F., 2008, MNRAS, 387, 536

\bibitem[{{Gibbons}, {Belokurov} \& {Evans}(2014){Gibbons}, {Belokurov}, \&
  {Evans}}]{Gibbons2014}
{Gibbons} S.~L.~J., {Belokurov} V., {Evans} N.~W., 2014, MNRAS, 445, 3788

\bibitem[{{Grand} {et~al}\mbox{.}(2019){Grand}, {Deason}, {White}, {Simpson},
  {G{\'o}mez}, {Marinacci}, \& {Pakmor}}]{Grand2019}
{Grand} R. J.~J., {Deason} A.~J., {White} S. D.~M., {Simpson} C.~M.,
  {G{\'o}mez} F.~A., {Marinacci} F., {Pakmor} R., 2019, MNRAS, 487, L72

\bibitem[{{Hammer} {et~al}\mbox{.}(2020){Hammer}, {Yang}, {Arenou}, {Wang},
  {Li}, {Bonifacio}, \& {Babusiaux}}]{Hammer2020}
{Hammer} F., {Yang} Y., {Arenou} F., {Wang} J., {Li} H., {Bonifacio} P.,
  {Babusiaux} C., 2020, ApJ, 892, 3

\bibitem[{{Hammer} {et~al}\mbox{.}(2015){Hammer}, {Yang}, {Flores}, {Puech}, \&
  {Fouquet}}]{Hammer2015}
{Hammer} F., {Yang} Y.~B., {Flores} H., {Puech} M., {Fouquet} S., 2015, ApJ,
  813, 110

\bibitem[{{Harris} \& {Canterna}(1979)}]{Harris1979}
{Harris} W.~E., {Canterna} R., 1979, ApJL, 231, L19

\bibitem[{{Hattori}, {Valluri} \& {Vasiliev}(2021){Hattori}, {Valluri}, \&
  {Vasiliev}}]{Hattori2021}
{Hattori} K., {Valluri} M., {Vasiliev} E., 2021, MNRAS, 508, 5468

\bibitem[{{Helmi} {et~al}\mbox{.}(2018){Helmi}, {Babusiaux}, {Koppelman},
  {Massari}, {Veljanoski}, \& {Brown}}]{Helmi2018}
{Helmi} A., {Babusiaux} C., {Koppelman} H.~H., {Massari} D., {Veljanoski} J.,
  {Brown} A. G.~A., 2018, Nature, 563, 85

\bibitem[{{Hopkins}(2015)}]{Hopkins2015}
{Hopkins} P.~F., 2015, MNRAS, 450, 53

\bibitem[{{Hopkins} {et~al}\mbox{.}(2018){Hopkins}, {Wetzel}, {Kere{\v{s}}},
  {Faucher-Gigu{\`e}re}, {Quataert}, {Boylan-Kolchin}, {Murray}, {Hayward},
  {Garrison-Kimmel}, {Hummels}, {Feldmann}, {Torrey}, {Ma},
  {Angl{\'e}s-Alc{\'a}zar}, {Su}, {Orr}, {Schmitz}, {Escala}, {Sanderson},
  {Grudi{\'c}}, {Hafen}, {Kim}, {Fitts}, {Bullock}, {Wheeler}, {Chan},
  {Elbert}, \& {Narayanan}}]{Hopkins2018}
{Hopkins} P.~F. {et~al.}, 2018, MNRAS, 480, 800

\bibitem[{{Jiao} {et~al}\mbox{.}(2021){Jiao}, {Hammer}, {Wang}, \&
  {Yang}}]{Jiao2021}
{Jiao} Y., {Hammer} F., {Wang} J.~L., {Yang} Y.~B., 2021, A\&A, 654, A25

\bibitem[{{Jing} \& {Suto}(2002)}]{Jing2002}
{Jing} Y.~P., {Suto} Y., 2002, ApJ, 574, 538

\bibitem[{{Kafle} {et~al}\mbox{.}(2012){Kafle}, {Sharma}, {Lewis}, \&
  {Bland-Hawthorn}}]{Kafle2012}
{Kafle} P.~R., {Sharma} S., {Lewis} G.~F., {Bland-Hawthorn} J., 2012, ApJ, 761,
  98

\bibitem[{{Kafle} {et~al}\mbox{.}(2014){Kafle}, {Sharma}, {Lewis}, \&
  {Bland-Hawthorn}}]{Kafle2014}
{Kafle} P.~R., {Sharma} S., {Lewis} G.~F., {Bland-Hawthorn} J., 2014, ApJ, 794,
  59

\bibitem[{{Kallivayalil} {et~al}\mbox{.}(2013){Kallivayalil}, {van der Marel},
  {Besla}, {Anderson}, \& {Alcock}}]{Kallivayalil2013}
{Kallivayalil} N., {van der Marel} R.~P., {Besla} G., {Anderson} J., {Alcock}
  C., 2013, ApJ, 764, 161

\bibitem[{{Karukes} {et~al}\mbox{.}(2020){Karukes}, {Benito}, {Iocco},
  {Trotta}, \& {Geringer-Sameth}}]{Karukes2020}
{Karukes} E.~V., {Benito} M., {Iocco} F., {Trotta} R., {Geringer-Sameth} A.,
  2020, JCAP, 2020, 033

\bibitem[{{Koposov} {et~al}\mbox{.}(2019){Koposov}, {Belokurov}, {Li}, {Mateu},
  {Erkal}, {Grillmair}, {Hendel}, {Price-Whelan}, {Laporte}, {Hawkins}, {Sohn},
  {del Pino}, {Evans}, {Slater}, {Kallivayalil}, {Navarro}, \& {Orphan Aspen
  Treasury Collaboration}}]{Koposov2019}
{Koposov} S.~E. {et~al.}, 2019, MNRAS, 485, 4726

\bibitem[{{Kuijken} \& {Gilmore}(1991)}]{Kuijken1991}
{Kuijken} K., {Gilmore} G., 1991, ApJL, 367, L9

\bibitem[{{K{\"u}pper} {et~al}\mbox{.}(2015){K{\"u}pper}, {Balbinot}, {Bonaca},
  {Johnston}, {Hogg}, {Kroupa}, \& {Santiago}}]{Kupper2015}
{K{\"u}pper} A. H.~W., {Balbinot} E., {Bonaca} A., {Johnston} K.~V., {Hogg}
  D.~W., {Kroupa} P., {Santiago} B.~X., 2015, ApJ, 803, 80

\bibitem[{{Lee} {et~al}\mbox{.}(2011){Lee}, {Beers}, {An}, {Ivezi{\'c}},
  {Just}, {Rockosi}, {Morrison}, {Johnson}, {Sch{\"o}nrich}, {Bird}, {Yanny},
  {Harding}, \& {Rocha-Pinto}}]{Lee2011}
{Lee} Y.~S. {et~al.}, 2011, ApJ, 738, 187

\bibitem[{{Li} {et~al}\mbox{.}(2021){Li}, {Hammer}, {Babusiaux}, {Pawlowski},
  {Yang}, {Arenou}, {Du}, \& {Wang}}]{Li2021}
{Li} H., {Hammer} F., {Babusiaux} C., {Pawlowski} M.~S., {Yang} Y., {Arenou}
  F., {Du} C., {Wang} J., 2021, arXiv e-prints, arXiv:2104.03974

\bibitem[{{Li} {et~al}\mbox{.}(2020){Li}, {Qian}, {Han}, {Li}, {Wang}, \&
  {Jing}}]{Li2020}
{Li} Z.-Z., {Qian} Y.-Z., {Han} J., {Li} T.~S., {Wang} W., {Jing} Y.~P., 2020,
  ApJ, 894, 10

\bibitem[{{Lindegren} {et~al}\mbox{.}(2018){Lindegren}, {Hern{\'a}ndez},
  {Bombrun}, {Klioner}, {Bastian}, {Ramos-Lerate}, {de Torres},
  {Steidelm{\"u}ller}, {Stephenson}, {Hobbs}, {Lammers}, {Biermann}, {Geyer},
  {Hilger}, {Michalik}, {Stampa}, {McMillan}, {Casta{\~n}eda}, {Clotet},
  {Comoretto}, {Davidson}, {Fabricius}, {Gracia}, {Hambly}, {Hutton}, {Mora},
  {Portell}, {van Leeuwen}, {Abbas}, {Abreu}, {Altmann}, {Andrei}, {Anglada},
  {Balaguer-N{\'u}{\~n}ez}, {Barache}, {Becciani}, {Bertone}, {Bianchi},
  {Bouquillon}, {Bourda}, {Br{\"u}semeister}, {Bucciarelli}, {Busonero},
  {Buzzi}, {Cancelliere}, {Carlucci}, {Charlot}, {Cheek}, {Crosta}, {Crowley},
  {de Bruijne}, {de Felice}, {Drimmel}, {Esquej}, {Fienga}, {Fraile}, {Gai},
  {Garralda}, {Gonz{\'a}lez-Vidal}, {Guerra}, {Hauser}, {Hofmann}, {Holl},
  {Jordan}, {Lattanzi}, {Lenhardt}, {Liao}, {Licata}, {Lister}, {L{\"o}ffler},
  {Marchant}, {Martin-Fleitas}, {Messineo}, {Mignard}, {Morbidelli}, {Poggio},
  {Riva}, {Rowell}, {Salguero}, {Sarasso}, {Sciacca}, {Siddiqui}, {Smart},
  {Spagna}, {Steele}, {Taris}, {Torra}, {van Elteren}, {van Reeven}, \&
  {Vecchiato}}]{Lindegren2018}
{Lindegren} L. {et~al.}, 2018, A\&A, 616, A2

\bibitem[{{Lindegren} {et~al}\mbox{.}(2021){Lindegren}, {Klioner},
  {Hern{\'a}ndez}, {Bombrun}, {Ramos-Lerate}, {Steidelm{\"u}ller}, {Bastian},
  {Biermann}, {de Torres}, {Gerlach}, {Geyer}, {Hilger}, {Hobbs}, {Lammers},
  {McMillan}, {Stephenson}, {Casta{\~n}eda}, {Davidson}, {Fabricius},
  {Gracia-Abril}, {Portell}, {Rowell}, {Teyssier}, {Torra}, {Bartolom{\'e}},
  {Clotet}, {Garralda}, {Gonz{\'a}lez-Vidal}, {Torra}, {Abbas}, {Altmann},
  {Anglada Varela}, {Balaguer-N{\'u}{\~n}ez}, {Balog}, {Barache}, {Becciani},
  {Bernet}, {Bertone}, {Bianchi}, {Bouquillon}, {Brown}, {Bucciarelli},
  {Busonero}, {Butkevich}, {Buzzi}, {Cancelliere}, {Carlucci}, {Charlot},
  {Cioni}, {Crosta}, {Crowley}, {del Peloso}, {del Pozo}, {Drimmel}, {Esquej},
  {Fienga}, {Fraile}, {Gai}, {Garcia-Reinaldos}, {Guerra}, {Hambly}, {Hauser},
  {Jan{\ss}en}, {Jordan}, {Kostrzewa-Rutkowska}, {Lattanzi}, {Liao}, {Licata},
  {Lister}, {L{\"o}ffler}, {Marchant}, {Masip}, {Mignard}, {Mints}, {Molina},
  {Mora}, {Morbidelli}, {Murphy}, {Pagani}, {Panuzzo}, {Pe{\~n}alosa Esteller},
  {Poggio}, {Re Fiorentin}, {Riva}, {Sagrist{\`a} Sell{\'e}s}, {Sanchez
  Gimenez}, {Sarasso}, {Sciacca}, {Siddiqui}, {Smart}, {Souami}, {Spagna},
  {Steele}, {Taris}, {Utrilla}, {van Reeven}, \& {Vecchiato}}]{Lindegren2020}
{Lindegren} L. {et~al.}, 2021, A\&A, 649, A2

\bibitem[{{Loebman} {et~al}\mbox{.}(2014){Loebman}, {Ivezi{\'c}}, {Quinn},
  {Bovy}, {Christensen}, {Juri{\'c}}, {Ro{\v{s}}kar}, {Brooks}, \&
  {Governato}}]{Loebman2014}
{Loebman} S.~R. {et~al.}, 2014, ApJ, 794, 151

\bibitem[{{Malhan} \& {Ibata}(2019)}]{Malhan2019}
{Malhan} K., {Ibata} R.~A., 2019, MNRAS, 486, 2995

\bibitem[{{Massari}, {Koppelman} \& {Helmi}(2019){Massari}, {Koppelman}, \&
  {Helmi}}]{Massari2019}
{Massari} D., {Koppelman} H.~H., {Helmi} A., 2019, A\&A, 630, L4

\bibitem[{{McMillan}(2017)}]{McMillan2017}
{McMillan} P.~J., 2017, MNRAS, 465, 76

\bibitem[{{McMillan} \& {Binney}(2013)}]{McMillan2013}
{McMillan} P.~J., {Binney} J.~J., 2013, MNRAS, 433, 1411

\bibitem[{{Monari} {et~al}\mbox{.}(2018){Monari}, {Famaey}, {Carrillo},
  {Piffl}, {Steinmetz}, {Wyse}, {Anders}, {Chiappini}, \&
  {Jan{\ss}en}}]{Monari2018}
{Monari} G. {et~al.}, 2018, A\&A, 616, L9

\bibitem[{{Moore} {et~al}\mbox{.}(1999){Moore}, {Ghigna}, {Governato}, {Lake},
  {Quinn}, {Stadel}, \& {Tozzi}}]{Moore1999}
{Moore} B., {Ghigna} S., {Governato} F., {Lake} G., {Quinn} T., {Stadel} J.,
  {Tozzi} P., 1999, ApJL, 524, L19

\bibitem[{{Mr{\'o}z} {et~al}\mbox{.}(2019){Mr{\'o}z}, {Udalski}, {Skowron},
  {Skowron}, {Soszy{\'n}ski}, {Pietrukowicz}, {Szyma{\'n}ski}, {Poleski},
  {Koz{\l}owski}, \& {Ulaczyk}}]{Mroz2019}
{Mr{\'o}z} P. {et~al.}, 2019, ApJL, 870, L10

\bibitem[{{Myeong} {et~al}\mbox{.}(2019){Myeong}, {Vasiliev}, {Iorio}, {Evans},
  \& {Belokurov}}]{Myeong2019}
{Myeong} G.~C., {Vasiliev} E., {Iorio} G., {Evans} N.~W., {Belokurov} V., 2019,
  MNRAS, 488, 1235

\bibitem[{{Naidu} {et~al}\mbox{.}(2021){Naidu}, {Conroy}, {Bonaca}, {Zaritsky},
  {Weinberger}, {Ting}, {Caldwell}, {Tacchella}, {Han}, {Speagle}, \&
  {Cargile}}]{Naidu2021}
{Naidu} R.~P. {et~al.}, 2021, arXiv e-prints, arXiv:2103.03251

\bibitem[{{Navarro} {et~al}\mbox{.}(2004){Navarro}, {Hayashi}, {Power},
  {Jenkins}, {Frenk}, {White}, {Springel}, {Stadel}, \& {Quinn}}]{Navarro2004}
{Navarro} J.~F. {et~al.}, 2004, MNRAS, 349, 1039

\bibitem[{{Necib} \& {Lin}(2021)}]{Necib2021}
{Necib} L., {Lin} T., 2021, arXiv e-prints, arXiv:2102.02211

\bibitem[{{Petersen} \& {Pe{\~n}arrubia}(2021)}]{Petersen2020}
{Petersen} M.~S., {Pe{\~n}arrubia} J., 2021, Nature Astronomy, 5, 251

\bibitem[{{Piffl} {et~al}\mbox{.}(2014){Piffl}, {Binney}, {McMillan},
  {Steinmetz}, {Helmi}, {Wyse}, {Bienaym{\'e}}, {Bland-Hawthorn}, {Freeman},
  {Gibson}, {Gilmore}, {Grebel}, {Kordopatis}, {Navarro}, {Parker}, {Reid},
  {Seabroke}, {Siebert}, {Watson}, \& {Zwitter}}]{Piffl2014}
{Piffl} T. {et~al.}, 2014, MNRAS, 445, 3133

\bibitem[{{Posti} {et~al}\mbox{.}(2015){Posti}, {Binney}, {Nipoti}, \&
  {Ciotti}}]{Posti2015}
{Posti} L., {Binney} J., {Nipoti} C., {Ciotti} L., 2015, MNRAS, 447, 3060

\bibitem[{{Posti} \& {Helmi}(2019)}]{Posti2019}
{Posti} L., {Helmi} A., 2019, A\&A, 621, A56

\bibitem[{{Pouliasis}, {Di Matteo} \& {Haywood}(2017){Pouliasis}, {Di Matteo},
  \& {Haywood}}]{Pouliasis2017}
{Pouliasis} E., {Di Matteo} P., {Haywood} M., 2017, A\&A, 598, A66

\bibitem[{{Read}(2014)}]{Read2014}
{Read} J.~I., 2014, Journal of Physics G Nuclear Physics, 41, 063101

\bibitem[{{Riello} {et~al}\mbox{.}(2021){Riello}, {De Angeli}, {Evans},
  {Montegriffo}, {Carrasco}, {Busso}, {Palaversa}, {Burgess}, {Diener},
  {Davidson}, {Rowell}, {Fabricius}, {Jordi}, {Bellazzini}, {Pancino},
  {Harrison}, {Cacciari}, {van Leeuwen}, {Hambly}, {Hodgkin}, {Osborne},
  {Altavilla}, {Barstow}, {Brown}, {Castellani}, {Cowell}, {De Luise},
  {Gilmore}, {Giuffrida}, {Hidalgo}, {Holland}, {Marinoni}, {Pagani},
  {Piersimoni}, {Pulone}, {Ragaini}, {Rainer}, {Richards}, {Sanna}, {Walton},
  {Weiler}, \& {Yoldas}}]{Riello2020}
{Riello} M. {et~al.}, 2021, A\&A, 649, A3

\bibitem[{{Riley} {et~al}\mbox{.}(2019){Riley}, {Fattahi}, {Pace}, {Strigari},
  {Frenk}, {G{\'o}mez}, {Grand}, {Marinacci}, {Navarro}, {Pakmor}, {Simpson},
  \& {White}}]{Riley2019}
{Riley} A.~H. {et~al.}, 2019, MNRAS, 486, 2679

\bibitem[{{Sofue}(2012)}]{Sofue2012}
{Sofue} Y., 2012, PASJ, 64, 75

\bibitem[{{Vasiliev}(2019{\natexlab{a}})}]{Vasiliev2019b}
{Vasiliev} E., 2019{\natexlab{a}}, MNRAS, 482, 1525

\bibitem[{{Vasiliev}(2019{\natexlab{b}})}]{Vasiliev2019a}
{Vasiliev} E., 2019{\natexlab{b}}, MNRAS, 484, 2832

\bibitem[{{Vasiliev}(2019{\natexlab{c}})}]{Vasiliev2019c}
{Vasiliev} E., 2019{\natexlab{c}}, MNRAS, 489, 623

\bibitem[{{Vasiliev} \& {Baumgardt}(2021)}]{Vasiliev2021b}
{Vasiliev} E., {Baumgardt} H., 2021, MNRAS, 505, 5978

\bibitem[{{Vasiliev}, {Belokurov} \& {Erkal}(2021){Vasiliev}, {Belokurov}, \&
  {Erkal}}]{Vasiliev2021a}
{Vasiliev} E., {Belokurov} V., {Erkal} D., 2021, MNRAS, 501, 2279

\bibitem[{{Voggel} {et~al}\mbox{.}(2016){Voggel}, {Hilker}, {Baumgardt},
  {Collins}, {Grebel}, {Husemann}, {Richtler}, \& {Frank}}]{Voggel2016}
{Voggel} K., {Hilker} M., {Baumgardt} H., {Collins} M. L.~M., {Grebel} E.~K.,
  {Husemann} B., {Richtler} T., {Frank} M.~J., 2016, MNRAS, 460, 3384

\bibitem[{{Wang} {et~al}\mbox{.}(2012){Wang}, {Frenk}, {Navarro}, {Gao}, \&
  {Sawala}}]{Wang2012}
{Wang} J., {Frenk} C.~S., {Navarro} J.~F., {Gao} L., {Sawala} T., 2012, MNRAS,
  424, 2715

\bibitem[{{Wang} {et~al}\mbox{.}(2019){Wang}, {Hammer}, {Yang}, {Ripepi},
  {Cioni}, {Puech}, \& {Flores}}]{Wang2019}
{Wang} J., {Hammer} F., {Yang} Y., {Ripepi} V., {Cioni} M.-R.~L., {Puech} M.,
  {Flores} H., 2019, MNRAS, 486, 5907

\bibitem[{{Wang} {et~al}\mbox{.}(2020){Wang}, {Han}, {Cautun}, {Li}, \&
  {Ishigaki}}]{Wang2020}
{Wang} W., {Han} J., {Cautun} M., {Li} Z., {Ishigaki} M.~N., 2020, Science
  China Physics, Mechanics, and Astronomy, 63, 109801

\bibitem[{{Watkins} {et~al}\mbox{.}(2019){Watkins}, {van der Marel}, {Sohn}, \&
  {Evans}}]{Watkins2019}
{Watkins} L.~L., {van der Marel} R.~P., {Sohn} S.~T., {Evans} N.~W., 2019, ApJ,
  873, 118

\bibitem[{{Wegg}, {Gerhard} \& {Bieth}(2019){Wegg}, {Gerhard}, \&
  {Bieth}}]{Wegg2019}
{Wegg} C., {Gerhard} O., {Bieth} M., 2019, MNRAS, 485, 3296

\bibitem[{{Wetzel} {et~al}\mbox{.}(2016){Wetzel}, {Hopkins}, {Kim},
  {Faucher-Gigu{\`e}re}, {Kere{\v{s}}}, \& {Quataert}}]{Wetzel2016}
{Wetzel} A.~R., {Hopkins} P.~F., {Kim} J.-h., {Faucher-Gigu{\`e}re} C.-A.,
  {Kere{\v{s}}} D., {Quataert} E., 2016, ApJL, 827, L23

\bibitem[{{York} {et~al}\mbox{.}(2000){York}, {Adelman}, {Anderson},
  {Anderson}, {Annis}, {Bahcall}, {Bakken}, {Barkhouser}, {Bastian}, {Berman},
  {Boroski}, {Bracker}, {Briegel}, {Briggs}, {Brinkmann}, {Brunner}, {Burles},
  {Carey}, {Carr}, {Castander}, {Chen}, {Colestock}, {Connolly}, {Crocker},
  {Csabai}, {Czarapata}, {Davis}, {Doi}, {Dombeck}, {Eisenstein}, {Ellman},
  {Elms}, {Evans}, {Fan}, {Federwitz}, {Fiscelli}, {Friedman}, {Frieman},
  {Fukugita}, {Gillespie}, {Gunn}, {Gurbani}, {de Haas}, {Haldeman}, {Harris},
  {Hayes}, {Heckman}, {Hennessy}, {Hindsley}, {Holm}, {Holmgren}, {Huang},
  {Hull}, {Husby}, {Ichikawa}, {Ichikawa}, {Ivezi{\'c}}, {Kent}, {Kim},
  {Kinney}, {Klaene}, {Kleinman}, {Kleinman}, {Knapp}, {Korienek}, {Kron},
  {Kunszt}, {Lamb}, {Lee}, {Leger}, {Limmongkol}, {Lindenmeyer}, {Long},
  {Loomis}, {Loveday}, {Lucinio}, {Lupton}, {MacKinnon}, {Mannery}, {Mantsch},
  {Margon}, {McGehee}, {McKay}, {Meiksin}, {Merelli}, {Monet}, {Munn},
  {Narayanan}, {Nash}, {Neilsen}, {Neswold}, {Newberg}, {Nichol}, {Nicinski},
  {Nonino}, {Okada}, {Okamura}, {Ostriker}, {Owen}, {Pauls}, {Peoples},
  {Peterson}, {Petravick}, {Pier}, {Pope}, {Pordes}, {Prosapio},
  {Rechenmacher}, {Quinn}, {Richards}, {Richmond}, {Rivetta}, {Rockosi},
  {Ruthmansdorfer}, {Sandford}, {Schlegel}, {Schneider}, {Sekiguchi}, {Sergey},
  {Shimasaku}, {Siegmund}, {Smee}, {Smith}, {Snedden}, {Stone}, {Stoughton},
  {Strauss}, {Stubbs}, {SubbaRao}, {Szalay}, {Szapudi}, {Szokoly}, {Thakar},
  {Tremonti}, {Tucker}, {Uomoto}, {Vanden Berk}, {Vogeley}, {Waddell}, {Wang},
  {Watanabe}, {Weinberg}, {Yanny}, {Yasuda}, \& {SDSS
  Collaboration}}]{York2000}
{York} D.~G. {et~al.}, 2000, AJ, 120, 1579

\bibitem[{{Zhao} {et~al}\mbox{.}(2012){Zhao}, {Zhao}, {Chu}, {Jing}, \&
  {Deng}}]{Zhao2012}
{Zhao} G., {Zhao} Y.-H., {Chu} Y.-Q., {Jing} Y.-P., {Deng} L.-C., 2012,
  Research in Astronomy and Astrophysics, 12, 723

\bibitem[{{Zhao}(1996)}]{Zhao1996}
{Zhao} H., 1996, MNRAS, 278, 488

\bibitem[{{Zinn}(1985)}]{Zinn1985}
{Zinn} R., 1985, ApJ, 293, 424

\end{thebibliography}

\begin{appendix}

\section{Dynamical modelling on GCs with Einasto profile}

\begin{figure*}
\center
\includegraphics[scale=1.0]{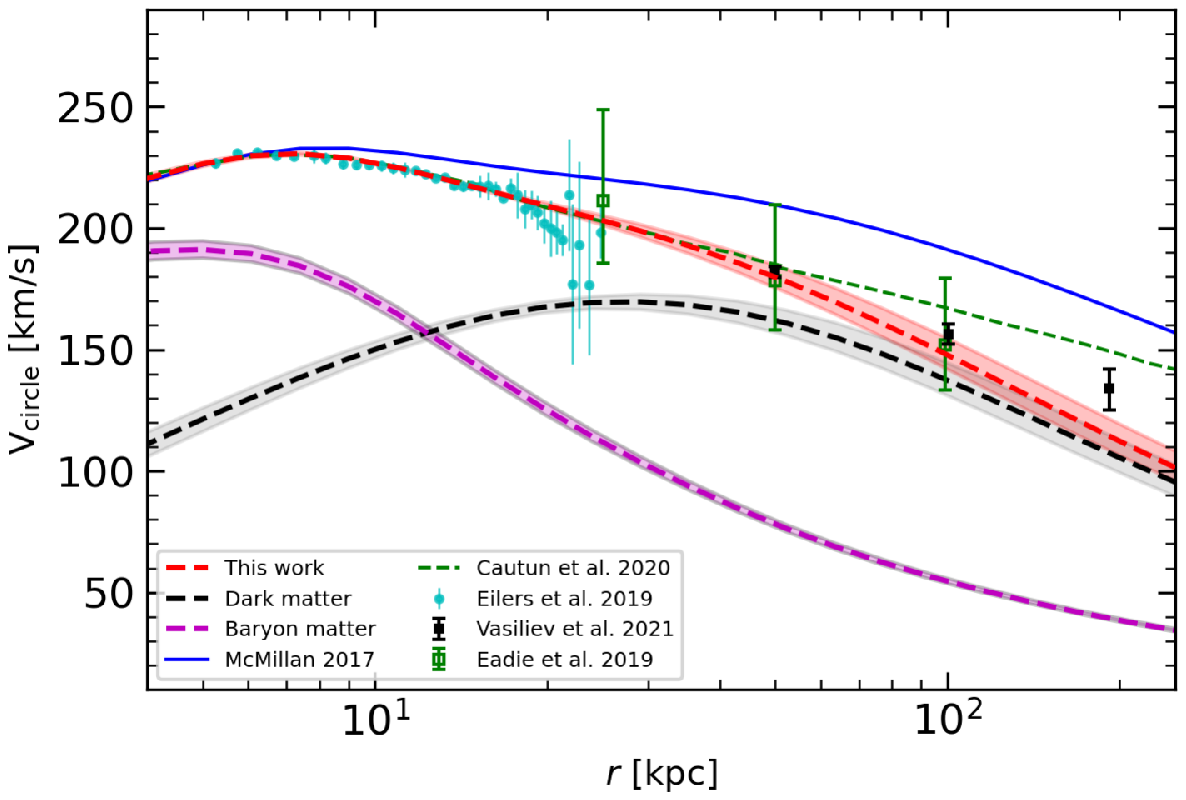}
\caption{It compares rotation curves derived from posterior distribution of our
models with literature. Here the Einasto dark matter profile is used. The shaded region indicate
the 68 percentile. }
\label{fig:RCEina}
\end{figure*}

\begin{figure*}
\center
\includegraphics[scale=1.2]{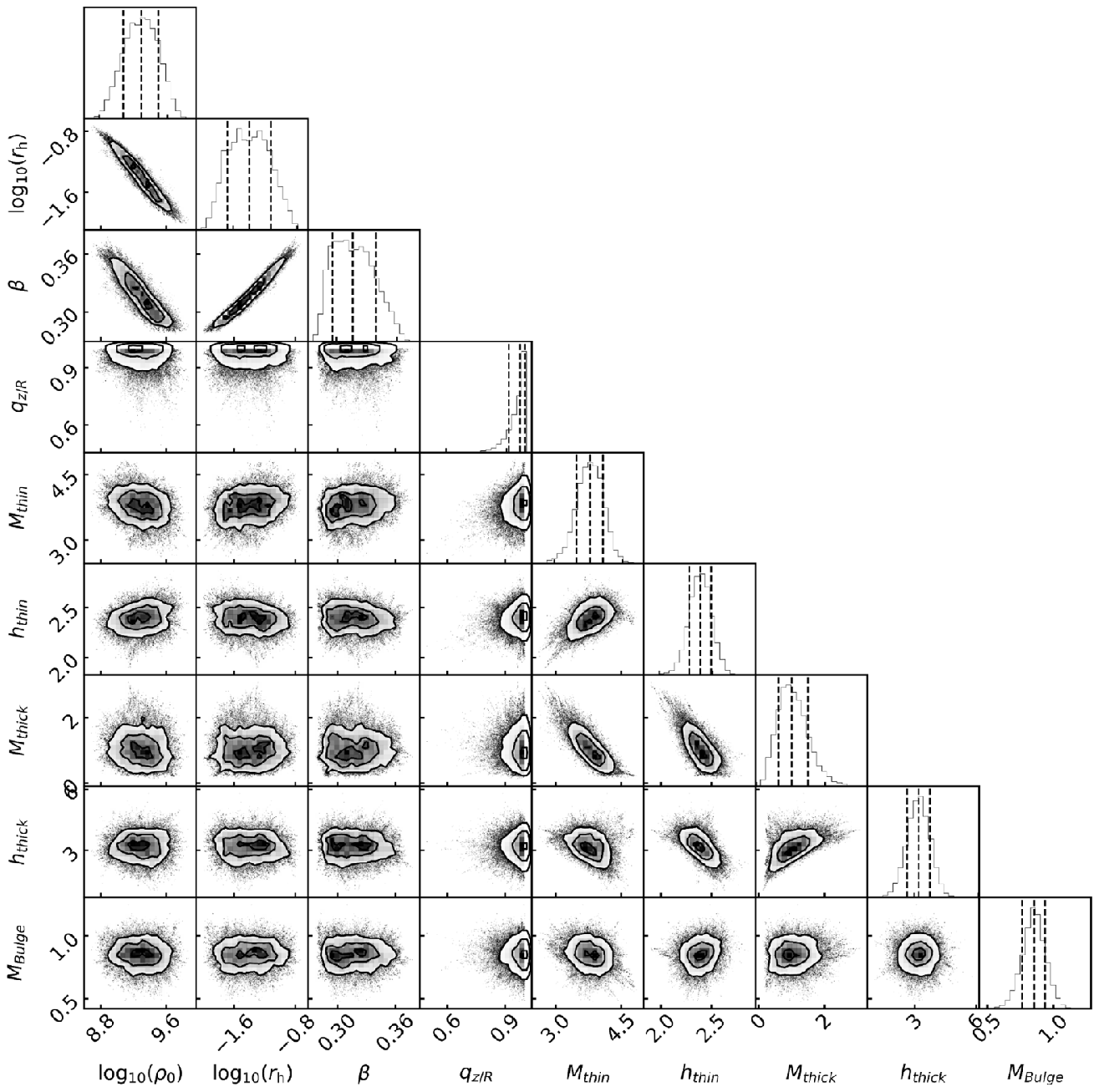}
\vspace*{5mm}
\caption{Posterior distribution of parameters for potential fields with model
using the Einasto's DM density profile (Eq.\ref{eq:Einasto}). The parameters $\log
\rho_0, \log r_\mathrm{h}, \beta$, are for parameters of
dark matter mass distribution. The parameters $M_\mathrm{thin},
M_\mathrm{thick}, M_\mathrm{bulge}$ are total mass for the thin and thick disk,
and bulge components, and their units is $10^{10}$ M$_{\odot}$. The parameters $h_\mathrm{thin}, h_\mathrm{thick}$
indicate the scale-length for thin and thick disk. 
The contour lines in each panel and 
the vertical lines in the marginal histograms are shown the 16\%, 50\%, 84\% percentiles.
}
\label{fig:Pot_Einasto}
\end{figure*}

\begin{figure*}
\center
\includegraphics[scale=1.2]{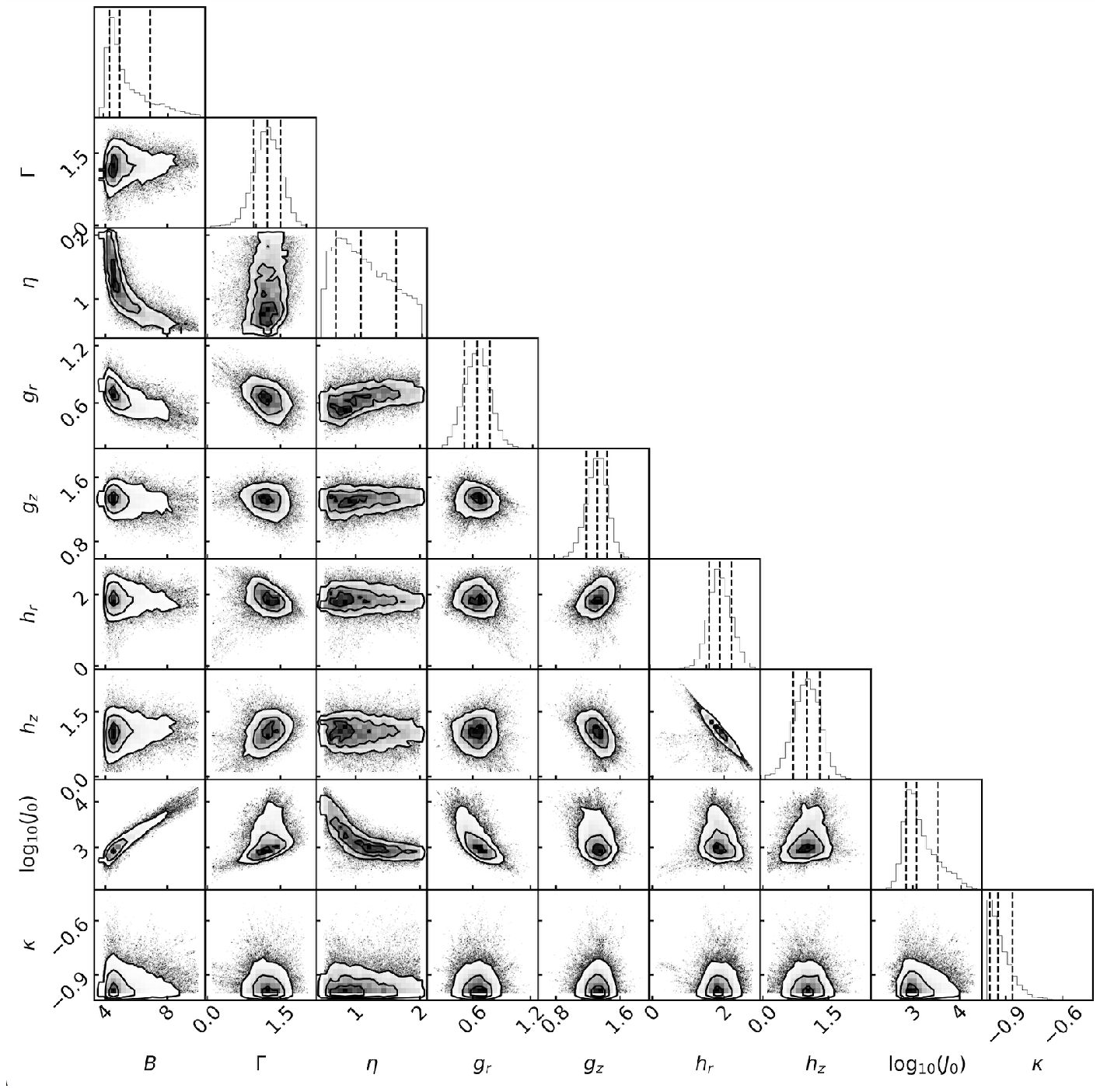}
\vspace*{5mm}
\caption{Posterior distribution of parameters of DF (eq. \ref{eq:DF}) of GCs
with model using the Einasto's DM density profile (Eq.\ref{eq:Einasto}).
The parameters $B$ and $\Gamma$ related to the outer and inner slopes of
DF, while $\eta$ determine the steepness of transition. The dimensionless
parameters $g_r, g_z, h_r, h_z$ control the density shape and the velocity
ellipsoid in the outer and inner region. The inner and outer regions are separated
with actions $\mathrm{J_0}$. $\kappa$ controls the rotation. The contour lines in each panel and
the vertical lines in the marginal histograms show the 16\%, 50\%, 84\% percentiles. 
}
\label{fig:DF_Einasto}
\end{figure*}

Figure~\ref{fig:RCEina} shows the result of the dynamical modeling of the GC system
based on the Einasto DM profile (Eq.\ref{eq:Einasto}).  The Einasto DM profile results
in a lower mass estimate at $r>100$ kpc compared to that from Zhao's DM profile
(Figure~\ref{fig:RC}). 
Fig.\ref{fig:Pot_Einasto} and Fig.\ref{fig:DF_Einasto} show the posterior
distribution for gravitational and DF parameters with Einasto's DM profile. 

We also notice that there are two GCs, Crater and Pyxis,  for which properties are still disputed.
It is not fully clear whether Crater is a dwarf or a GC
\citep{Bonifacio2015,Voggel2016}, and \citet{Fritz2017} argued that Pyxis is
accreted from a disrupted dwarf. We tested our results by excluding the two GCs
from our samples, and found the MW total mass are
$6.77_{-1.74}^{+3.00}\times10^{11}$M$_{\odot}$ and $5.36_{-0.68}^{+0.81}\times
10^{11}$ M$_{\odot}$ with Zhao and Einasto profile, respectively.  These values
are lower than that derived with the full samples (Table \ref{tab:FitPars}).
We note that the multi-population in the samples have no significant effect on
our results as being test with our FIRE2 simulation. The proper motion of Pyxis
in the $Gaia$ DR2 and EDR3 \citep[this work and][]{Vasiliev2019a,Vasiliev2021b}
is smaller than that used in \citet{Fritz2017}, which corresponds to
velocity decrease by $\sim 70$ km/s and making Pyxis bound to the MW system.
Therefore, including it in the samples is reasonable.

\section{Dynamical modelling on GCs with data from Vasiliev et al. 2021b} \label{sec:RCVasiliev}

\begin{figure*}
\center
\includegraphics[scale=1.0]{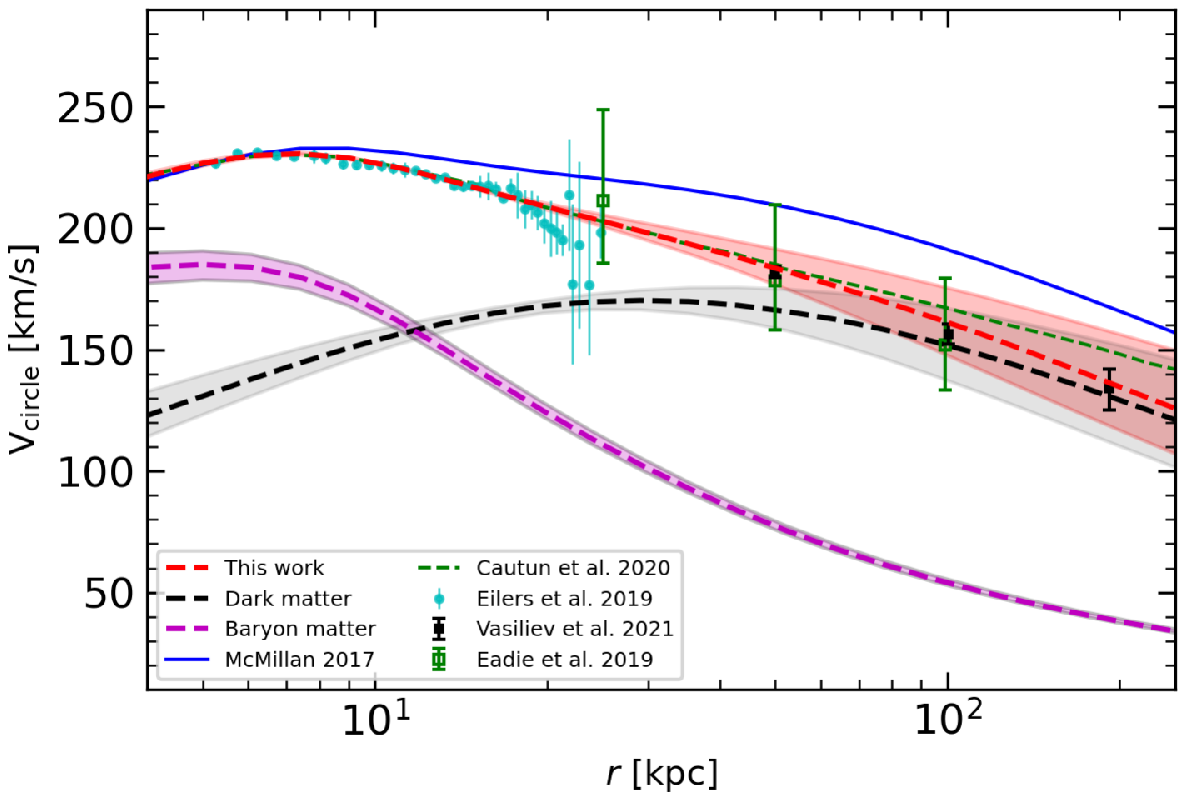}
\caption{Comparing rotation curves derived from posterior distribution of our
models with literature with GCs data from \citet{Vasiliev2021b}. Zhao's dark matter profile are used. 
The shaded region indicate the 68 percentile. }
\label{fig:RCVasiliev}
\end{figure*}

In order to check how the results changing with PM results of \citet{Vasiliev2021b}, we 
also run our code on GCs data from \citet{Vasiliev2021b} as shown in Fig.\ref{fig:RCVasiliev}.
The results is very similar to that with GC PMs derived in this work. The total mass of MW with 
the data of \citet{Vasiliev2021b} is $8.27_{-2.14}^{+3.59} \times 10^{11} $ M$_{\odot}$, which 
is slightly larger than the result with our GCsdata, but still within the error bars.

\end{appendix}

\end{document}